\documentclass{article}

\author{Hao Chen\\
Department of Mathematics\\
Zhongshan University\\
Guangzhou,Guangdong 510275\\
People's Republic of China\\
e-mail: mcsch@zsu.edu.cn}
\title{\bf Quantum Entanglement and Geometry of Determinantal Varieties}
\date{April, 2003}

\begin{document}

\maketitle
\begin{abstract}
Quantum entanglement was first recognized as a feature of quantum mechanics in the famous paper of Einstein, Podolsky 
and Rosen [18]. Recently it has been realized that quantum entanglement is a key ingredient in quantum computation, 
quantum communication and quantum cryptography ([16],[17],[6]). In this paper, we introduce algebraic sets, which are 
determinantal varieties in the complex projective spaces or the products of complex projective spaces, for the mixed 
states in  bipartite or multipartite quantum systems as their invariants under local unitary transformations. These 
invariants are naturally arised from the physical consideration of measuring mixed states by separable pure states. In 
this way algebraic geometry and complex differential geometry of these algebraic sets turn to be powerful tools for 
the understanding of quantum enatanglement. Our construction has applications in the following important topics in 
quantum information theory: 1) separability criterion, it is proved the algebraic sets have to be the sum of the  
linear subspaces if the mixed states are separable; 2) lower bound of Schmidt numbers, that is, generic low rank  
bipartite mixed states are entangled in many degrees of freedom; 3) simulation of Hamiltonians, it is proved the 
simulation of semi-positive Hamiltonians of the same rank implies the projective isomorphisms of the corresponding 
algebraic sets; 4) construction of bound enatanglement, examples of the entangled mixed states which are invariant 
under partial transpositions (thus PPT bound entanglement) are constructed systematically from our new separability 
criterion. On the other hand many examples of entangled mixed states with rich algebraic-geometric  structure in their 
associated determinantal varieties are constructed and studied from this point of view.

\end{abstract}

\section*{1.Introduction}

A bipartite pure quantum state $|\psi \rangle \in H=H_A^m \otimes H_B^n$, where $H_A, H_B$ are finite dimensional 
Hilbert spaces and the tensor inner product is used on $H$, is called entangled if it cannot be written as $|\psi 
\rangle =|\psi_A \rangle \otimes |\psi_B \rangle$ for some $|\psi_A \rangle \in H_A$ and $|\psi_B \rangle \in H_B$. A 
mixed state (or a density matrix) $\rho$, which is a positive semidefinite operator on $H$ with trace 1, is called 
entangled if it cannot be written as\\

$$
\begin{array}{ccccc}
\rho=\Sigma_i p_i P_{|\psi_i^A\rangle} \otimes P_{|\psi_i^B \rangle}
\end{array}
(1)
$$
, for some set of states $|\psi_i^A \rangle \in H_A, |\psi_i^B \rangle \in H_B$ and $p_i \geq 0$. Here $P_v$ for a 
state (unit vector) $v$ means the (rank 1) projection operator to the vector $v$.  If the mixed state $\rho$ can be 
written in the form of (1), it is called separable (see [28],[39][42]).\\

For the mixed state $\rho$ on $H=H_A \otimes H_B$ ,we have the following partial transposition $\rho^{PT}$ and the 
partial trace $tr_B(\rho)$ on $H_A^m$($tr_A(\rho)$ on $H_B^n$ can be defined similarly) defined as follows.\\

$$
\begin{array}{cccccc}
\langle ij|\rho^{PT}|kl \rangle = \langle il|\rho|kj \rangle\\
\langle i|tr_B(\rho)|k \rangle =\Sigma_j  \langle ij|\rho|kj \rangle
\end{array}
(2)
$$
, where $\{|1 \rangle,...,|m \rangle \},\{|1 \rangle ,...,|n \rangle \},\{|11 \rangle ,...|1n \rangle ,...|m1 
\rangle,...,|mn \rangle \}$ are the standard orthogonal basis of $H_A,H_B,H$ respectively. It is easy to check that 
the definition is not dependent on the special basis ([28],[39],[42]).\\

For multipartite quantum systems, there are similar definitions of entangled and separable mixed states (see 
[28],[39],[42]). We restrict ourselves to the case of the tripartite case. Let $\rho$ be a mixed state in $H=H_A 
\otimes H_B \otimes H_C$. $\rho$ is called separable if it can be written as \\

$$
\begin{array}{ccccc}
\rho=\Sigma_i p_i P_{|\psi_i^A \rangle} \otimes P_{|\psi_i^B \rangle} \otimes P_{|\psi_i^C \rangle}
\end{array}
(3)
$$
, for some set of states $|\psi_i^A \rangle \in H_A, |\psi_i^B \rangle  \in H_B,|\psi_i^C \rangle \in H_C$ and $p_i 
\geq 0$. If the mixed state $\rho$ cannot be written in the form of (3), it is called entangled. Sometimes we also 
consider the separability relative to the cut A:BC (B:AC etc.), that means $\rho$ is considered as a mixed state in 
the bipartite quantum system $H=H_A \otimes (H_B \otimes H_C)$.\\

For a n-party quantum system $H=H_{A_1}^{m_1} \otimes ...\otimes H_{A_n}^{m_n}$, local unitary transformations  
(acting on a mixed state $\rho$ by $U \rho U^{\dagger}$, where $\dagger$ is the adjoint.) are those unitary 
transformations of the form $U=U_{A_1} \otimes ... \otimes U_{A_n}$, where $U_{A_i}$ is a unitary transformation on 
$H_{A_i}^{m_i}$ for $i=1,...,n$. We can check that all eigenvalues (spectra) of $\rho$ (global spectra) and 
$,tr_{A_{i_1}...A_{i_l}}(\rho)$ of mixed states $\rho$, where $i_1,...,i_l \in \{1,...,n\}$, (local spectra) are 
invariant under local unitary transformations, and the invariants in the examples of [33] and [34] are more or less 
spectra-involved. It is clear that separability (or being entangled) is an invariant property under local unitary 
transformations. For a mixed state $\rho$ , to judge whether it is entangled or separable and decide its entangled 
class (i.e. the equivalent class of all entangled (or separable) mixed states which are equivalent to $\rho$ by local 
unitary transformations) is a fundamental problem in the study of quantum entanglement ([28],[42]).Thus for the 
purpose to quantify entanglement, any good measure of entanglement must be invariant under local transformations 
([33],[34],[28],[42]). Another important concept is the distillable mixed state, which means that some singlets can be 
extracted from it by local operations and classical communication (LOCC) (see [28]).  A mixed state $\rho$ in $H$ is 
distillable if and only if for some  $t$, there exists projections $P_A: H_A^{\otimes t} \rightarrow H^2$ and $P_B: 
H_B^{\otimes t} \rightarrow H^2$, where $H^2$ is a 2 dimensional Hilbert space, such that the mixed state $(P_A 
\otimes P_B)\rho^{\otimes t}(P_A \otimes P_B)^{\dagger}$ is a entanlged state in $H^2 \otimes H^2$ (see [28]). A mixed 
state which cannot be distilled is called bounded entangled mixed state.\\

The phonomenon of quantum entanglement lies at the heart of quantum mechanics since the famous Einstein, Podolsky and 
Rosen [18] paper (see [5],[28],[39]). Its importance lies not only in philosophical consideration of the nature of 
quantum theory, but also in applications where it has emerged recently that quantum entanglement is the key ingredient 
in quantum computation [16] and communication [6] and plays an important role in cryptography ([17]).These new 
applications of quantum entanglement have stimulated tremendous studies of quantum entanglements of both pure and 
mixed states from both theoretical and experimental view, for surveys we refer to [5],[28],[32],[39] and [42].\\

To find good necessary condition of separability (separability criterion) is a fundamental problem in the study of 
quantum entanglement ([28],[42]). Bell-type  inequalities ([28]) and entropy criterion ([28]) are well-known numerical 
criterion of separable states. In 1996, Peres [38] gave a striking simple criterion which asserts that a separable 
mixed state $\rho$ necessarily has (semi) positive partial transposition (PPT), which has been proved by Horodeckis 
([24]) also a sufficient condition of separability in $ 2 \times 2 $ and $2 \times 3$ systems. The significance of PPT 
property is also reflected in the facts that PPT mixed states satisfy Bell inequalities ([44]) and cannot be distilled 
([28],[26]), thus the first several examples of the PPT entangled mixed states ([25]) indicated that the new 
phenomenon that there is bound entanglement in nature ([26]). These examples were constructed from the so-called range 
criterion of P.Horodecki (see [25],[28]). However constructing PPT entangled mixed states (thus bound entanglement) is 
exceedingly difficult task ([7]), and the only known systematic way of such construction is the context of 
unextendible product base (UPB) in [7], which works in both bipartite and multipartite case and is also based on 
P.Horodecki's range criterion. The most recent disorder criterion of separability  in [37], which is stronger than 
entropy criterion, was proved by the mathematics of majorization.\\

It has been realized that the entanglement of tripartite pure quantum states is not a trivial extension of the 
entanglement of bipartite pure quantum states ([21]). Recently C.H.Bennett, etc., [8] studied the exact and asymptotic 
entanglement measure of multipartite pure states, which showed  essential difference to that of  bipartite pure 
states. On the other hand Acin, etc., [2] proved a generalization of Schmidt decomposition for pure triqubit states, 
which seems impossible to be generalized to arbitrary multipartite case . Basically, the understanding of multipartite 
quantum entanglement for both pure and mixed states, is much less advanced.\\

It is clear that any separability criterion for bipartite mixed states, such as Peres PPT criterion [38] and Horodecki 
range criterion ([25],[28]), can be applied to multipartite mixed states for their separability under various cuts. 
For example, from Peres PPT criterion, a separable multipartite mixed state necessarily has all its partial 
transpositions semi-positive. In [27], Horodeckis studied the separability criterion of multipartite mixed states by 
linear maps. Classification of triqubit mixed states inspired by Acin, etc., [2] was studied in [3].\\ 

There have been many interesting works [11],[12],[31],[35] and [43] for understanding quantum entanglemment and 
related problems from the view of Representation Theory and Topology.\\ 

The physical motivation of this paper is as follows. We consider the following situation. Alice and Bob share a 
bipartite quantum system $H_A^m \otimes H_B^n$, and they have a mixed state $\rho$. Now they want to understand the 
entanglement properties of $\rho$. It is certain that they can prepare any separable pure state  $|\phi_1 \rangle 
\otimes |\phi_2 \rangle$ seaprately. Now they measure $\rho$ with this separable pure state, the expectation value is 
$ \langle \phi_1 \otimes \phi_2|\rho|\phi_1 \otimes \phi_2 \rangle$. If Alice's pure state $|\phi_1 \rangle$ is fixed, 
then $ \langle \phi_1 \otimes \phi_2|\rho|\phi_1 \otimes \phi_2 \rangle $ is a Hermitian bilinear form on Bob's pure 
states (i.e., on $H_B^n$). We denote this bilinear form by $ \langle \phi_1|\rho|\phi_1 \rangle$. Intuitively the {\em 
degenerating locus} $V_A^k(\rho)=\{|\phi_1 \rangle \in P(H_A^m): rank (\langle \phi_1|\rho|\phi_1 \rangle) \leq k\}$ , 
where $P(H_A^m)$ is the projective space of all pure states in $H_A^m$, should contain the physical information of 
$\rho$ and it is almost obvious that these {\em degenerating locus} are invariant under local unitary transformations. 
In a multipartite quantum system, a similar consideration leads to some Hermitian bilinear forms on some of its parts 
and similarly we can consider the {\em degenerating locus} of these Hermitian bilinear forms. We prove that these {\em 
degenerating locus} are  algebraic sets, which are determinantal varieties, for the mixed states in both bipartite and 
multipartite quantum systems. They have the following properties:\\

1) When we apply local unitary transformations on the mixed state the corresponding  algebraic sets are  changed by 
local (unitary) linear transformations, and thus these invariants can be used to distinguish inequivalent mixed states 
under local unitary transformations;\\

2) The algebraic sets have to be  linear (the sum of some linear subspaces) if the mixed state is separable, and thus 
we give a new separability criterion;\\

3) The algebraic sets are independent of eigenvalues and only measure the positions of eigenvectors of the mixed 
states;\\

4) These algebraic sets can be calculated easily.\\

From our construction here, we establish a connection between  Quantum Entanglement and  Algebraic Geometry. Actually 
from our results below, we can see that if the Fubini-Study metric of the projective complex space is used, the metric 
properties of these algebraic sets are also preserved when local  unitary transformations are applied on the mixed 
state. Hence we establish a connection between Quantum Entanglement and both the Algebraic Geometry and complex 
Differential  Geometry of these algebraic sets. Any algebraic-geometric or complex differential geometric invariant of 
the algebraic set of the mixed state is an invariant of the mixed state under local unitary transformations.\\

The determinantal varieties ([23] Lecture 9 and [4] Cha.II) have been studied from different motivations such as  
geometry of curves ([4],[19]), geometry of determinantal varieties([20]), Hodge theory ([22]), commutative algebra 
([15]) and even combinatorics ([1]). It is interesting to see that it can be useful even in quantum information 
theory. We refer to the books [23] and [4] for the standard facts about determinantal varieties.\\

The paper is organized as follows. We define the algebraic sets of the mixed states and prove their basic properties 
including the separability criteria based on these algebraic sets in section 2. In section 3, we indicate briefly how 
numerical invariants of the bipartite or multipartite mixed states under local unitary transformations can be derived 
from these algebraic sets. As an easiest example, Schmidt rank of a pure state in bipartite quantum systems, a 
classical concept in quantum entanglement, is showed to be the codimention of the algebraic set. From our invariants, 
a LOWER bound of Schmidt numbers of mixed states (defined in [41]) is given (Theorem 4), which implies that the 
``generic'' rank $m$ mixed states on $H_A^m \otimes H_B^m$ have relatively high Schmidt numbers. Many examples of 
entangled mixed states corresponding to the famous determinantal varieties in algebraic geometry, such as Segre 
varieties, rational normal scrolls and generic determinantal varieties are constructed in section 4. We also show that 
a well-known theorem of D.Eisenbud ([20]) can help us to construct many entangled bipartite mixed states of low ranks. 
In section 5 we introduce the generalized Smolin states, which is a natural extension of Smolin's physical 
construction [40] from algebraic-geometric view. In section 6 and 7, it is proved that these algebraic sets are 
nonempty for low rank mixed states, and indicate how a {\em finer} result with the same idea (Theorem 9') can be 
potentially used to treat the entanglement properties of high rank mixed states. Based on the algebraic sets 
introduced in section 2, a necessary condition about the simulating Hamiltonians efficiently using local unitary 
transformations is given in section 8. In section 9, we give a continuous family of bipartite mixed states, tripartite 
pure states and bipartite Hamiltonians with the property that the eigenvalues (spectra) of them and their partial 
traces are constant, however their entanglement properties are distinct. This offers strong evidences that it is
 hopeless to characterize the entanglement properties by only using eigenvalue spectra. In section 10, we illustrate 
by an explicit example that our separability crieterion can be used to construct PPT entangled mixed states (thus 
bound entanglement) systematically.\\

\section*{2. Invariants and separability criteria}

Now we use the coordinate form of the physical consideration of {\em degenerating locus} described in the 
introduction. Let $H=H_{A}^m \otimes H_{B}^n$,  $\{|ij \rangle\}$, where, $i=1,...,m$ and $j=1,...,n$, be the standard 
basis and $\rho$ be a mixed state on $H$. We represent the matrix of $\rho$ in the basis $\{|11\rangle,...|1n 
\rangle,...,|m1 \rangle,...,|mn\rangle\}$, and consider $\rho$ as a blocked matrix $\rho=(\rho_{ij})_{1 \leq i \leq m, 
1 \leq j \leq m}$ with each block $\rho_{ij}$ a $n \times n$ matrix corresponding to the $|i1 \rangle,...,|in \rangle 
$ rows and the $|j1 \rangle,...,|jn \rangle$ columns. For any pure state $|\phi_1 \rangle=r_1|1 \rangle+...+r_m|m 
\rangle \in P(H_A^m)$ the matrix of the Hermitian linear form $\langle\phi_1|\rho|\phi_1 \rangle$ with the basis $|1 
\rangle ,...,|n \rangle$ is $\Sigma_{i,j} r_i r_j^{\dagger} \rho_{ij}$. Thus the {\em degenerating locus} are as 
follows.\\

{\bf Definition 1.}{\em  We define 

$$
\begin{array}{ccccc}
V_{A}^k(\rho)=\{(r_1,...,r_m)\in CP^{m-1}:rank( \Sigma_{i,j}r_i r_j^{\dagger} \rho_{ij}) \leq k \}
\end{array}
(4)
$$
where $k=0,1,...,n-1$.Similarly $V_{B}^k(\rho) \subseteq CP^{n-1}$, where $k=0,1,...,m-1$, can be defined.}\\

{\bf Example 1.} Let $\rho=\frac{1}{mn}I_{mn}$, the maximally mixed state, we easily have 
$V_{A}^k(\rho)=\{(r_1,...,r_n):rank(\Sigma r_i r_i^{\dagger}I_n) \leq k\}=\emptyset$ for $k=0,1,...,n-1$.\\

{\bf Example 2.} Let $H=H_A^2 \otimes H_B^n$, $T_1,T_2$ be 2 $n \times n$ matrices of rank $n-1$ such that the $n 
\times (2n)$ matrix $(T_1,T_2)$ has rank $n$. Let $T'$ be a $2n \times 2n$ matrix with 11 block $T_1$, 22 block 
$T_2$,12 and 21 blocks 0. Its rows correspond to the standard base $|11 \rangle,...,|1n \rangle,|21\rangle,...,|2n 
\rangle$ of $H$. Let $\rho=\frac{1}{D}TT^{\dagger}$ (where $D$ is a normalizing constant) be a mixed state on $H$. It 
is easy to check that $\rho$ is of  rank $2n-2$ and $V_A^{n-1}(\rho)=\{(r_1,r_2):r_1r_2=0\}$.\\

Let $H=H_{A}^m \otimes H_{B}^n \otimes H_{C}^l$. The standard orthogonal basis is of $H$ is $|ijk\rangle$, where, 
$i=1,...,m$,$j=1,...,n$ and $k=1,...,l$, and $\rho$ is a mixed state on $H$. We represent the matrix of $\rho$ in the 
base $\{|111 \rangle,...|11l\rangle,...,|mn1\rangle,...,|mnl\rangle\}$ as
 $\rho=(\rho_{ij,i'j'})_{1 \leq i ,i' \leq m, 1 \leq j ,j' \leq n}$, and  
$\rho_{ij,i'j'}$ is a $l \times l$ matrix. Consider $H$ as a bipartite system as $H=(H_{A}^m \otimes H_{B}^n) \otimes 
H_{C}^l$, then we have $V_{AB}^k(\rho)=\{(r_{11},...,r_{mn}) \in CP^{mn-1}:rank( \Sigma r_{ij}r_{i'j'}^{\dagger} 
\rho_{ij,i'j'}) \leq k\}$ defined as above. This set is actually the {\em degenerating locus}  of the Hermitian 
bilinear form $\langle\phi_{12}|\rho|\phi_{12}\rangle$ on $H_C^l$ for the given pure state $|\phi_{12} \rangle 
=\Sigma_{i,j}^{m,n} r_{ij} |ij \rangle \in P(H_A^m \otimes H_B^n)$. When the finer cut A:B:C is considered, it is 
natural to take $|\phi_{12} \rangle$ as a separable pure state $|\phi_{12} \rangle =|\phi_1 \rangle \otimes |\phi_2 
\rangle$, ie., there exist $|\phi_1 \rangle =\Sigma_i r_i^1 |i \rangle \in P(H_A^m),|\phi_2 \rangle=\Sigma_j r_j^2 |j 
\rangle \in P(H_B^n)$ such that $r_{ij}=r_i^1r_j^2$. In this way the tripartite mixed state $\rho$ is measured by 
tripartite separable pure states $|\phi_1 \rangle \otimes |\phi_2 \rangle \otimes  |\phi_3 \rangle$. Thus it is 
natural we define $V_{A:B}^k(\rho)$ as follows. It is the {\em degenerating locus}  of the bilinear form $\langle 
\phi_1 \otimes \phi_2 |\rho|\phi_1 \otimes \phi_2 \rangle$ on $H_C^l$.\\ 

{\bf Definition 2.} {\em  Let $\phi: CP^{m-1} \times CP^{n-1} \rightarrow CP^{mn-1}$ be the mapping defined by\\

$$\begin{array}{ccccccccc}
\phi(r_1^1,...r_m^1,r_1^2,...,r_n^2)=(r_1^1r_1^2,...,r_i^1r_j^2,...r_m^1r_n^2)
\end{array}
(5)
$$

(i.e., $r_{ij}=r_i^1 r_j^2$ is introduced.)

Then $V_{A:B}^k(\rho)$ is defined as the preimage $\phi^{-1}(V_{AB}^k(\rho))$.}\\

Similarly $V_{B:C}^k(\rho),V_{A:C}^k(\rho)$ can be defined. In the following statement we just state the result for 
$V_{A:B}^k(\rho)$. The conclusion holds similarly for other $V's$.\\

For the mixed state $\rho$ in the multipartite system $H=H_{A_1}^{m_1} \otimes \cdots \otimes H_{A_k}^{m_k}$, we want 
to study the entanglement under the cut $ A_{i_1}:A_{i_2}:...:A_{i_l}:(A_{j_1}...A_{j_{k-l}})$, where $\{i_1,...,i_l\} 
\cup \{j_1,...j_{k-l}\}=\{1,...k\}$. We can define the set $V_{A_{i_1}:...:A_{i_l}}^k(\rho)$ similarly. \\

{\bf Theorem 1.} {\em  Let $T=U_{A} \otimes U_{B}$, where $U_{A}$ and $U_{B}$ are the unitary transformations on 
$H_{A}^m$ and $H_{B}^n$ respectively. Then $V_{A}^k(T(\rho))=U_{A}^{-1}(V_{A}^k(\rho))$, that is $V_{A}^k(\rho)$ 
(resp. $V_{B}^k(\rho)$) is an ``invariant'' up to a linear (unitary if Fubuni-Study metric is used) transformation of 
$CP^{m-1}$ of the mixed state $\rho$ under local unitary transformations.}\\

{\bf Proof.} Let $U_{A}=(u_{ij}^{A})_{1 \leq i \leq m,1 \leq j \leq m}$ and $U_{B}=(u_{ij}^{B})_{1 \leq i \leq n, 1 
\leq j \leq n}$ be the matrices in the standard orthogonal basis. Then the matrix of $T(\rho)$ under the standard 
orthogonal base $\{|11 \rangle,...,|1n  \rangle,...,|m1 \rangle,...,|mn \rangle\}$, as a blocked matrix, is 
$T(\rho)=(\Sigma_{l,h} u_{il}^{A} U_{B}(\rho_{lh})U_{B}^{\dagger}(u_{jh}^{A})^{\dagger})_{1\leq i \leq m, 1 \leq j 
\leq m}$. Hence \\

$$
\begin{array}{cccccccc}
V_{A}^k(T(\rho))=\{(r_1,...,r_m): rank (\Sigma _{l,h} (\Sigma_i r_i u_{il}^{A})(\Sigma_j r_j 
u_{jh}^{A})^{\dagger}U_{B}(\rho_{lh})U_{B}^{\dagger}) \leq k \}

\end{array}
(6)
$$

We set $r_l'=\Sigma_i r_i u_{il}^{A}$ for $l=1,...,m$.Then\\

$$
\begin{array}{ccccc} 
\Sigma _{l,h} (\Sigma_i r_i u_{il}^{A})(\Sigma_j r_j u_{jh}^{A})^{\dagger}U_{B}(\rho_{lh})U_{B}^{\dagger}\\
=U_{B}(\Sigma_{lh} r_l' (r_h')^{\dagger} (\rho_{lh}))U_{B}^{\dagger} 

\end{array}
(7)
$$

Thus\\
$$
\begin{array}{ccccccc}
V_A^k(T(\rho))=\{(r_1,...,r_m): rank (\Sigma_{lh} r_l' (r_h')^{\dagger} (\rho_{lh}))\leq k \}

\end{array}
(8)
$$

and our conclusion follows. \\

{\bf Theorem 1'.} {\em  Let $T=U_{A} \otimes U_{B} \otimes U_{C}$, where $U_{A},U_{B}$ and $U_{C}$ are unitary 
transformations on $H_{A}^m, H_{B}^n$ and $H^l$ respectively. Then $V_{A:B}^k(T(\rho))=U_{A}^{-1} \times 
U_{B}^{-1}(V_{A:B}^k(\rho))$,  that is $V_{A:B}^k(\rho)$ is an ``invariant'' up to a linear (unitary if product 
Fubini-Study metric is used) transformation of $CP^{m-1} \times CP^{n-1}$  of the mixed state $\rho$ under local 
unitary transformations.}\\

{\bf Proof.} Let $U_{A}=(u_{ij}^{A})_{1 \leq i \leq m,1 \leq j \leq m}$, $U_{B}=(u_{ij}^{B})_{1 \leq i \leq n, 1 \leq 
j \leq n}$ and $U_{C}=(u_{ij}^{C})_{1 \leq i \leq l, 1 \leq j \leq l}$, be the matrix in the standard orthogonal 
basis.\\

Recall the proof of Theorem 1, we have $V_{AB}(T(\rho))=(U_{A} \otimes U_{B})^{-1}(V_{AB}(\rho))$ under the coordinate 
change\\

$$
\begin{array}{cccccccc} 
r_{kw}'=\Sigma_{ij} r_{ij} u_{ik}^{A} u_{jw}^{B}\\
=\Sigma_{ij}r_i^1 r_j^2 u_{ik}^{A} u_{jw}^{B}\\
=\Sigma_{ij} (r_{i}^1 u_{ik}^A) (r_{j}^2 u_{jw}^B)\\
=(\Sigma_i r_i^1 u_{ik}^A)(\Sigma_j r_j^2 u_{jw}^B)
\end{array}
(9)
$$

for $k=1,...,m,w=1,...,n$. Thus our conclusion follows from the definition.\\

Similarly, we have the following result in general case.\\

{\bf Theorem 1''.} {\em  Let $T=U_{A_{i_1}} \otimes \cdots \otimes U_{A_{i_l}} \otimes U_{j_1...j_{k-l}}$, where 
$U_{A_{i_1}},...,U_{A_{i_l}}, U_{j_1...j_{k-l}}$ are unitary transformations on 
$H_{A_{i_1}}^{m_{i_1}},...,H_{A_{i_l}}^{m_{i_l}}$  and $(H_{A_{j_1}}^{m_{j_1}} \otimes ...\otimes 
H_{A_{j_{k-l}}}^{m_{j_{k-l}}})$ respectively. Then $V_{A_{i_1}:...:A_{i_l}}^k(T(\rho))=U_{A_{i_1}}^{-1} \times \cdots 
\times U_{A_{i_l}}^{-1}(V_{A_{i_1}:...:A_{i_l}}^k(\rho))$.}\\

{\bf Remark 1.} Since $U_{A_{i_1}}^{-1} \times \cdots \times  U_{A_{i_l}}^{-1}$ certainly preserves the standard 
Euclid metric of complex linear space and hence the (product) Fubini-Study metric of the product of projective complex 
spaces, all metric properties of $V_{A_{i_1}:...:A_{i_l}}^k(\rho)$ are preserved when the local unitary 
transformations are applied to the mixed state $\rho$.\\

Moreover from the proof it is easy to see that all algebraic-geometric properties (since $V_A^k(\rho),V_B^k(\rho)$ are 
algebraic sets as proved in Theorem 2) of $V_A^k(\rho)$ (resp. $V_B^k(\rho)$) are preserved even under local linear 
inversible transformations (i.e., $T=T_A \otimes T_B$ where $T_A,T_B$ are just linear inversible operators of 
$H_A^m,H_B^n$.)\\

We observe $V_A^k(\rho^{PT})=(V_A^k(\rho))^{\dagger}$, here $\dagger$ is the conjugate mapping of $CP^{m-1}$ defined 
by $(r_1,...,r_m)^{\dagger}=(r_1^{\dagger},...,r_m^{\dagger})$. It is clear that this property holds for other $V$'s 
invariants.\\

{\bf Theorem 2.} {\em $V_{A}^k(\rho)$ (resp. $V_{B}^k(\rho)$) is an algebraic set in $CP^{m-1}$ (resp. $CP^{n-1}$).}\\

From Definition 2 and Theorem 2 we immediately have the following result.\\

{\bf Theorem 2'.} {\em $V_{A:B}^k(\rho)$ is an algebraic set in $CP^{m-1} \times CP^{n-1}$.}\\

The general result can be stated as follows.\\

{\bf Theorem 2''.} {\em $V_{A_{i_1}:...:A_{i_l}}^k(\rho)$ is an algebraic set in $CP^{m_{i_1}-1} \times 
CP^{m_{i_l}-1}$.}\\

It is easy to see from Definitions that we just need to prove Theorem 2.\\

{\bf Theorem 3.} {\em If $\rho$ is a separable mixed state, $V_{A}^k(\rho)$ (resp. $V_{B}^k(\rho)$) is a linear subset 
in $CP^{m-1}$ (resp. $CP^{n-1}$),i.e., it is the sum of the linear subspaces.}\\

In the following statement we give the separability criterion of the mixed state $\rho$ under the cut A:B:C. The 
``linear subspace of $CP^{m-1} \times CP^{n-1}$'' means the product of a linear subspace in $CP^{m-1}$ and a linear 
subspace in $CP^{n-1}$.\\

{\bf Theorem 3'.} {\em If $\rho$ is a separable mixed state on $H=H_{A}^m \otimes H_{B}^{n} \otimes H_{C}^l$ under the 
cut A:B:C, $V_{A:B}^k(\rho)$ is a linear subset in $CP^{m-1} \times CP^{n-1}$, i.e., it is the sum of the linear 
subspaces.}\\

The general result can be stated as follows.\\

{\bf Theorem 3''.} {\em If $\rho$ is a separable mixed state on $H=H_{A_1}^{m_1} \otimes \cdots \otimes H_{A_k}^{m_k}$ 
under the cut $ A_{i_1}:A_{i_2}:...:A_{i_l}:(A_{j_1}...A_{j_{k-l}})$, $V_{A_{i_1}:...:A_{i_l}}^k(\rho)$ is a linear 
subset in $CP^{m_{i_1}-1}  \times \cdots  \times CP^{m_{i_l}-1}$,i.e., it is the sum  of the linear subspaces.}\\

We just prove Theorem 3 and Theorem 3'. The proof of Theorem 3'' is similar.\\

For the purpose to prove Theorem 2 and 3 we need the following lemmas.\\

{\bf Lemma 1.} {\em Let $\{e_1,...,e_h\}$ be the orthogonal basis of a $h$ dimension Hilbert space $H$, 
$\rho=\Sigma_{l=1}^{t} p_l P_{v_l}$, where $v_l$ is unit vector in $H$ for $l=1,...,t$ ,   $v_l=\Sigma_{k=1}^{h} 
a_{kl} e_k$ and $A=(a_{kl})_{1\leq k \leq h, 1 \leq l \leq t}$ is the $h \times t$ matrix. Then the matrix of $\rho$ 
with the base $\{e_1,...,e_h\}$ is $APA^{\dagger}$, where $P$ is the diagonal matrix with diagonal entries 
$p_1,...,p_h$.}\\

{\bf Proof}. We note that the matrix of $P_{v_l}$ with the basis is $\alpha_l \alpha_l^{\dagger}$ where 
$\alpha_l=(a_{1l},...,a_{hl})^{\tau}$ is just the expansion of $v_l$ with the basis. The conclusion follows 
immediately.\\

The following conclusion is a direct matrix computation from Lemma 1 or see [25].\\

{\bf Corollary 1.} {\em Suppose $p_i>0$, then the image of $\rho$ is the linear span of vectors $v_1,...,v_t$.}\\

Now let $H$ be the $H_{A}^m \otimes H_{B}^n$ ,$\{e_1,...,e_{mn}\}$ be the standard orthogonal base 
$\{|11\rangle,...,|1n\angle,...,|m1\rangle,...,|mn\rangle \}$ and $\rho= \Sigma_{l=1}^{t} p_l P_{v_l}$ with positive 
$p_l$'s be a mixed state on $H$. We may consider the $mn\times t$ matrix $A$ as a $m\times 1$ blocked matrix with each 
block $A_w$, where $w=1,...,m$, a $n\times t$ matrix corresponding to $\{|w1\rangle,...,|wn\rangle\}$. Then it is easy 
to see $\rho_{ij}=A_i P A_j^{\dagger}$, where $i=1,...m,j=1,...,m$. Thus\\

$$
\begin{array}{cccccc}
\Sigma r_ir_j^{\dagger} \rho_{ij}=(\Sigma r_i A_i)P(\Sigma r_i  A_i)^{\dagger}
\end{array}
(10)
$$

{\bf Lemma 2.} {\em $\Sigma r_i r_j^{\dagger} \rho_{ij}$ is a (semi) positive definite $n\times n$ matrix. Its rank 
equals to the rank of $(\Sigma r_i A_i)$.}\\

{\bf Proof.} The first conclusion is clear. The matrix $\Sigma r_i r_j^{\dagger} \rho_{ij}$ is of rank $k$ if and only 
if there exist $n-k$ linear independent vectors $c^j=(c_1^j,...,c_n^j)$ with the property.\\

$$
\begin{array}{cccccccccc}
c^j(\Sigma_{ij}  r_ir_j^{\dagger} \rho_{ij})(c^{j})^{\dagger}=\\
(\Sigma_i r_i c^jA_i)P(\Sigma_i r_i c^{j}A_i)^{\dagger}=0
\end{array}
(11)
$$

Since $P$ is a strictly positive definite matrix, our conclusion follows immediately.\\

{\bf Proof of Theorem 2.} From Lemma 1 , we know that $V_{A}^k(\rho)$ is the zero locus of all $(k+1)\times (k+1)$ 
submatrices of $(\Sigma r_iA_i)$. The conclusion is proved.\\

Because the determinants of all $(k+1) \times (k+1)$ submatrices of $(\Sigma r_i A_i)$ are homogeneous polynomials of 
degree $k+1$ , thus $V_{A}^k (\rho)$(resp. $V_{B}^k(\rho)$) is an algebraic subset (called determinantal varieties in 
algebraic geometry [18],[20]) in $CP^{m-1}$(resp. $CP^{n-1}$).\\

The point here is: for different representations of $\rho$ as $\rho=\Sigma_j p_j P_{v_j}$ with $p_j$'s  positive real 
numbers, the determinantal varieties from their corresponding $\Sigma_i r_iA_i$'s are the same.\\

Now suppose that the mixed state $\rho$ is separable, i.e  there are unit product vectors $a_1 \otimes b_1,....,a_s 
\otimes b_s$ such that $\rho=\Sigma_{l=1}^{s}q_l P_{a_l \otimes b_l}$ , where $q_1,...q_s$ are positive real numbers. 
Suppose $a_u=a_u^1 |1 \rangle+...+a_u^m |m \rangle,b_u= b_u^1 |1 \rangle+...+b_u^n|n \rangle$ for $u=1,...,s$. Hence 
the vector representation of $a_u \otimes b_u$ with the standard basis is $a_u \otimes b_u= \Sigma_{ij} a_u^ib_u^j |ij 
\rangle$. Consider the corresponding $mn \times s$ matrix $C$ of $a_1 \otimes b_1,...,a_s \otimes b_s$ as in Lemma 1, 
we have $\rho=CQC^{\dagger}$, where $Q$ is diagonal matrix with diagonal entries $q_1,...,q_s$. As before we consider 
$C$ as $m\times 1$ blocked matrix with blocks $C_w$, $w=1,...m$. Here $C_w$ is a $n \times s$ matrix of the form 
$C_w=(a_j^{w}b_j^i)_{1\leq i \leq n, 1 \leq j \leq s}=BT_w$ , where $B=(b_j^i)_{1\leq i \leq n, 1\leq j\leq s}$ is a 
$n \times s$ matrix and $T_w$ is a diagonal matrix with diagonal entries $a_1^{w},...,a_{s}^{w}$. Thus from Lemma 1, 
we have $\rho_{ij}=C_i Q C_j^{\dagger}=B(T_i Q T_j^{\dagger})B^{\dagger}=B T_{ij}B^{\dagger}$, where $T_{ij}$ is a 
diagonal matrix with diagonal entries $q_1 a_1^i(a_1^j)^{\dagger},...,q_s a_s^i(a_s^{j})^{\dagger}$.\\

{\bf Proof of Theorem 3.} As in the proof of Theorem 2, we have\\

$$
\begin{array}{ccccccc}
\Sigma r_ir_j^{\dagger} \rho_{ij}=\Sigma r_i r_j^{\dagger}BT_{ij}B^{\dagger}\\
=B(\Sigma r_ir_j^{\dagger}T_{ij})B^{\dagger}
\end{array}
(12)
$$

Here we note  $\Sigma r_i r_j^{\dagger}T_{ij}$ is a diagonal matrix with diagonal entries\\ 
$ q_1(\Sigma r_i a_1^{i})(\Sigma r_i a_1^{i})^{\dagger},...,q_s(\Sigma r_i a_s^{i})(\Sigma r_i a_s^{i})^{\dagger}$. 
Thus $\Sigma r_ir_j^{*} \rho_{ij}=BGQG^{\dagger}B^{\dagger}$, where $G$ is a diagonal matrix with diagonal entries $ 
\Sigma r_ia_1^{i},...,\Sigma r_ia_s^{i}$. Because $Q$ is a strictly positive definite matrix, from Lemma 1 we know 
that  $\Sigma r_i r_j^{\dagger} \rho_{ij}$ is of rank smaller than $k+1$ if and only if the rank of $BG$ is strictly 
smaller than $k+1$. Note that $BG$ is just the multiplication of $s$ diagonal entries of $G$ (which are linear forms 
of $r_1,...,r_m$) on the $s$ columns of $B$,  thus the determinants of all $(k+1) \times (k+1)$ submatrices of $BG$ 
(in the case $ s \geq k+1$, otherwise automatically linear)are the multiplications of a constant (possibly zero) and 
$k+1$ linear forms of $r_1,...,r_m$. Thus the conclusion is proved.\\

From Lemma 1 and the proof of Theorem 2 and 3 , $\Sigma_i r_iA_i$ play a key role. If we take the standard $\rho= 
\Sigma_{j=1}^r p_j P_{\varphi_j}$, where $p_j , \varphi_j, j=1,...,r$ are eigenvalues and eigenvectors, the 
corresponding $\Sigma_i r_i A_i$ measures the {\em geometric positions} of eigenvectors in $H_A^m \otimes H_B^n$. It 
is obvious from the proof of Theorem 2, the non-local invariants defined in Definition 1 are independent of 
$p_1,...,p_r$ , the global eigenvalue spectra of the mixed states.\\

{\bf Proof of Theorem 3'.} We first consider the separability of $\rho$ under the cut AB:C, i.e., $\rho= 
\Sigma_{f=1}^g p_f P_{a_f \otimes b_f}$, where $a_f \in H_{A}^m \otimes H_{B}^n$ and $b_f \in H_{C}^l$ for 
$f=1,...,g$. Consider the separability of $\rho$ under the cut A:B:C, we have  $a_f=a_f' \otimes a_f''$ , $a_f' \in 
C_{A}^m, a_f'' \in C_{B}^m$. Let $a_f=(a_f^1,...,a_f^{mn}), a_f'=(a_f'^1,...,a_f'^m)$ and 
$a_f''=(a_f''^1,...,a_f''^n)$ be the coordinate forms with the standard orthogonal basis $\{|ij \rangle\}$, $\{|i 
\rangle\}$ and $\{|j \rangle\}$ respectively, we have that $a_f^{ij}=a_f'^i a_f''^j$. Recall the proof of Theorem 3, 
the diagonal entries of $G$ in the proof of Theorem 3 are\\

$$
\begin{array}{cccccccc}
\Sigma_{ij}r_{ij} a_f^{ij}=\\
\Sigma_{ij} r_i^1 a_f'^i r_j^2 a_f''^j=\\
(\Sigma_i r_i^1 a_f'^i)(\Sigma_j r_j^1 a_f''^j)
\end{array}
(13)
$$

Thus as argued in the proof of Theorem 3 , $V_{A:B}^k(\rho)$ has to be the zero locus of the multiplications of the 
linear forms in (13). The conclusion is proved.\\

\section*{3. Numerical invariants}

It is a standard fact in algebraic geometry that $V's$ defined in section 2 are the sum of irreducible algebraic 
varieties( components). Suppose $V_A^k(\rho)=V_1 \cup \cdots \cup V_t$. From Theorem 1 and Remark 1, we know that $t$ 
is a numerical invariant of $\rho$ when local linear inversible  transformations are applied to $\rho$. Actually, 
since there are a lot of numerical algebriac-geometric invariants of these components, e.g., dimensions, cohomology 
classes (represented by $V_i$'s in $H^{*}(CP^{m-1})$, cohomology rings of $V_i$'s,  etc. We can get many numerical 
invarints of the mixed state when local linear inversible transformations are applied to them. In this way, we get a 
very powerful tool of numerical invarints to distinguish the entangled classes of the mixed states in a composite 
quantum systems.\\

On the other hand, if local unitary transformations are applied to the mixed states, it is known that even the metric 
properties of $V's$ (the metric on $V$ is from the standard Fubini-Study metric of projective spaces) are invariant. 
Thus any complex differential geometric quantity, such as, the volumes of $V_i$'s, the integrations (over the whole 
component) of some curvature functions of $V_i$'s, are the invariants of the mixed states under local unitary 
transformations.\\

For any given pure state $|v \rangle$  in a bipartite quantum system, $|v \rangle \in H_A^m \otimes H_B^n$, there 
exist orthogonal basis $|\phi_1 \rangle,...,|\phi_m \rangle$ of $H_A^m$ and orthogonal basis $|\psi_1 
\rangle,...,|\psi_n \rangle$ of $H_B^n$ , such that, $|v \rangle=\lambda_1 |\phi_1 \rangle \otimes|\psi_1 \rangle 
+\cdots +\lambda_d |\phi_d \rangle \otimes|\psi_d \rangle$, where $d \leq min\{m,n\}$. This is Schmidt decomposition 
(see [39]). It is clear that $d$ is an invariant under local unitary transformations. This number is called the 
Schmidt rank of the pure state $|v \rangle$. It is clear that $|v \rangle$ is separable if and only if its Schmidt 
rank is 1. Schmidt rank of pure states in a bipartite quantum systems is a classical concept in the theory of quantum 
entanglement, it is actually the codimension of the invariant $V_A^0(\rho)$ for the pure state $\rho=P_{|v \rangle 
}$.\\

Let $\rho=P_{|v \rangle}$ be a pure state in $H_A^m \otimes H_B^n$ with $m \leq n$. From Theorem 1 about the 
invariance of $V_A^0(\rho)$, we can compute it from its Schmidt decomposition $|v \rangle= \Sigma_{i=1}^d \lambda_i 
|\phi_i \rangle \otimes |\psi_i \rangle$. It is clear that $V_A^0(\rho) =\{(r_1,...,r_m) \in CP^{m-1}: 
(\lambda_1r_1,...,\lambda_d r_d,0,...,0)^{\tau}=0\}$.\\

{\bf Proposition 1.} {\em For the pure state $\rho=P_{|v \rangle}$, $d=m$ if and only if $V_A^0(\rho)=\emptyset$ and 
$d=m-1-dim(V_A^0(\rho))=n-1-dim(V_B^0(\rho))$ if $d \leq m-1$.}\\

In this way we show that the Schmidt rank of a pure state is just the codimension of the algebraic set, and thus it 
seems interesting to study the quantity $m-1-dim(V_A^k(\rho))$ for mixed states, since it is non-local invariant and 
the generalization of the classical concept of Schmid rank of pure states.\\

In [41] the concept of  Schmidt numbers of mixed states was introduced as the minimum Schmidt rank of pure states that 
are needed to construct such mixed states. For a bipartite mixed state $\rho$, it has Schmidt number $k$ if and only 
if for any decomposition $\rho=\Sigma_i p_i P_{v_i}$ with positive real numbers $p_i$'s and pure states $v_i$'s, at 
least one of the pure states $v_i$'s has Schmidt rank at least $k$, and there exists such a decomposition with all 
pure states $v_i$'s Schmidt rank at most $k$. It is clear that $\rho$ is separable if and only if its Schmidt number 
is 1.\\

From our invariants, a LOWER bound of Schmidt numbers of mixed states is given as in the following Theorem 4, which 
implies that the generic rank $m$ mixed states on $H_A^m \otimes H_B^m$ have relatively high Schmidt numbers and thus 
entangled. This is an extension of our previous results in [14].\\

{\bf Theorem 4.} {\em Let $\rho$ be a mixed state on $H_A^m \otimes H_B^m$ of rank $r$ and Schmidt number $k$. Suppose 
$V_A^{m-t}(\rho)=\emptyset$, then $k \geq \frac{m}{r-m+t}$.}\\

{\bf Proof.} Take any representation $\rho= \Sigma_{i=1}^t p_i P_{v_i}$ with $p_i$'s positive,  and the maximal 
Schmidt rank of $v_i$'s is $k$. We observe that it is only needed to take $r$ linear independent vectors in 
$\{v_1,...,v_t\}$ to compute the rank of $\Sigma_i r_i A_i$ in Lemma 1, since the columns in $\Sigma_i r_i A_i$ 
corresponding to the remaining $t-r$ vectors are linear dependent on the columns corresponding to these $r$ linear 
independent vectors. For the purpose that the rank of these $r$ columns in $\Sigma_i r_i A_i$ is not bigger than 
$m-t$, we just need $r-m+t$ of these columns to be zero. On the other hand, from Proposition 1, the dimension of the 
linear subspace $(r_1,...,r_m) \in H_A^m$, such that the corresponding column of $v_i$ in $\Sigma_i r_i A_i$ is zero,  
is exactly $m-k(v_i)$, where $k(v_i)$ is the Schmidt rank of $v_i$. Thus we know that there is at least one nonzero 
$(r_1,...,r_m)$ such that $\Sigma_i r_i A_i$ is of rank smaller than $m-t+1$ if $ m >k(r-m+t)$. The conclusion is 
proved.\\

The physical implication of Theorem 4 is interesting. We apply it to  the rank $lt$ mixed states on $H_A^{lt} \otimes 
H_B^{lt}$ with $t \geq l$. From the Proposition in p.67 of [4] (or see Proposition 4 below),  $V_A^{(l-1)t}(\rho)$ of 
the ``generic'' rank $lt$ mixed states $\rho$'s has codimension $t^2>lt-1$ in $CP^{lt-1}$, thus empty. We know that 
the Schmidt numbers of these generic rank $lt$ mixed states are at least $l$. Hence we have the following result.\\

{\bf Corollary 2.} {\em For rank $n^2$ generic mixed states in a bipartite quantum system $H_A^{n^2} \otimes 
H_B^{n^2}$, their Schmidt numbers are greater than or equal to  $n$.}\\

For example, to construct  ``generic'' rank $9$ mixed states on $H_A^9 \otimes H_B^9$, pure states of Schmidt rank at 
least 3 have to be used.\\ 

For any pure state in a bipartite quantum system $H=H_A^m \otimes H_B^n$ , it can be written as a linear combination 
of at most $min\{m,n\}$ 2-way orthogonal separable pure states ([39],[8]) from Schmidt decomposition. For multipartite 
pure states, there is no direct generaliztion of Schmidt decomposition, and those multipartite pure states with  m-way 
orthogonal decompositions can be distilled to cat states (see [8]). From the results in [2], it is known that we need 
at least 5 terms of ``orthogonal''  separable pure states to write a generic pure state in $H_A^2 \otimes H_B^2 
\otimes H_C^2$ as a linear combination of them. This phenomenon is a remarkable difference between bipartite pure 
state entanglement and multipartite pure state entanglement. In the following statement it is showed  what happens for 
generic pure states in $H_A^{n^2} \otimes H_B^{n^2} \otimes H_C^{n^2}$.\\

{\bf Theorem 5.} {\em For a generic pure state $|\psi \rangle=\Sigma_{i=1}^{n^2} \lambda_i |\psi_i^{12} \rangle 
\otimes |\psi_i^3 \rangle$ in a tripartite quantum system $H_A^{n^2} \otimes H_B^{n^2} \otimes H_C^{n^2}$, where 
$|\psi_1^3 \rangle,...,|\psi_{n^2}^3 \rangle$ are mutually orthogonal unit vectors in $H_C^{n^2}$ and $|\psi_1^{12} 
\rangle,...,|\psi_{n^2}^{12} \rangle$ are pure states in $H_A^{n^2} \otimes H_B^{n^2}$, then there exists one of 
$|\psi_1^{12} \rangle,...,|\psi_{n^2}^{12}\rangle$ with Schmidt rank at least $n$.}\\

{\bf Proof.} It is clear that $tr_C(P_{|\psi \rangle} )=\Sigma_{i=1}^{n^2} |\lambda_i|^2 P_{|\psi_i^{12}\rangle}$ is a 
generic rank $n^2$ mixed state in $H_A^{n^2} \otimes H_B^{n^2}$. From Theorem 4 , at least one of $|\psi_1^{12} 
\rangle,...,|\psi_{n^2}^{12} \rangle$ has Schmidt rank (as a pure state in $H_A^{n^2} \otimes H_B^{n^2}$) at least 
$n$. Thus our conclusion is proved.\\

\section*{4. Examples}

Now we give some examples to show how to use Theorem 1,2,3 to construct and distinguish the entangled classes of the 
mixed states.\\

{\bf Example 3.} Let $H=H_A^m \otimes H_B^n$ and $\rho_{a_1,...,a_n}=\frac{1}{n}(\Sigma_{i=1}^n P_{a_i \otimes |i 
\rangle})$, where $a_i$, $i=1,...,n$, are unit vectors in $H_A^m$. This is a rank n separable mixed state. Suppose 
$a_i=(a_i^1,...,a_i^m)$, $i=1,...,n$,  the expansion with respect to the standard basis $|1 \rangle,...,|m \rangle$ of 
$H_A^m$. Let $l_i(r_1,...,r_m)=a_i^1 r_1+...+a_i^mr_m$ for $i=1,...,n$ be n linear forms. It is easy to check that 
$V_A^{n-1}(\rho)=\{(r_1,...,r_m): l_1 \cdots l_n=0 \}$.\\ 

{\bf Proposition 2.} {\em The mixed states $\rho_{a_1,...,a_n}$ and $\rho_{b_1,...,b_n}$ are equivalent under the 
local unitary transformations if and only if there exists a unitary transformation $U_A$ on $H_A^m$ such that the n 
vectors $b_1,..,b_n$ are exactly $U_A(a_1),...,U_A(a_n)$,i.e., $b_i=U_A(a_{i_j})$, where 
$\{i_1,...,i_n\}=\{1,...,n\}$.}\\

{\bf Proof.} The ``if'' part is clear. Let $l'_i(r_1,...,r_m)=b_i^1r_1+...+b_i^mr_m$ for $i=1,...,n$. Then 
$V_A^{n-1}(\rho_{a_1,...,a_n})$ (resp. $V_A^{n-1} (\rho_{b_1,...,b_n})$ )are the union of n hyperplanes defined by 
$l_i=0$ (resp. $l'_i=0$) for $i=1,...,n$. It should be noted here these hyperplances are counted by multiplicities. 
From Theorem 1 we get the conclusion.\\

Segre variety $\Sigma_{n,m}$ , which is the image of the following map, $\sigma: CP^n \times CP^m \rightarrow 
CP^{(n+1)(m+1)-1}$ where $\sigma ([X_0,...,X_n],[Y_0,...,Y_m])=[...,X_iY_j,...]$, is a famous determinantal variety 
(see [23], pp.25-26). It is clear that Segre variety is irreducible and not a linear subvariety. We consider the Segre 
variety $\Sigma_{1,m}$ in the case $n=1$, actually $\Sigma_{1,m}=\{(r_1,...,r_{2n}): rank(M) \leq 1\}$ where $M$ is 
the following matrix\\

$$
\left(
\begin{array}{cccccccc}
r_1&r_2&...&r_n\\
r_{n+1}&r_{n+2}&...&r_{2n}
\end{array}
\right)
(14)
$$

{\bf Example 4 (entangled mixed state from Segre variety).} Let $H=H_A^{2m} \otimes H_B^2$, $|\phi_i \rangle 
=\frac{1}{\sqrt{2}}(|i1 \rangle+|(m+i)2 \rangle)$ for $i=1,...,m$ and $\rho=\frac{1}{m}(P_{|\phi_1 
\rangle}+...+P_{|\phi_m \rangle})$. This is a mixed state of rank $m$. By computing $\Sigma_i r_i A_i$ as in the proof 
of Theorem 2, we get $V_A^1(\rho)=\Sigma_{1,m}$. Thus $\rho$ is an entangled mixed state.\\

The rank 1 locus $X_{l,n-l-1}=\{rank(R) \leq 1\}$ of the following $2 \times (n-1)$ matrix $R$\\

$$
\left(
\begin{array}{ccccccccccccccccccc}
r_0&...&r_{l-1}&r_{l+1}&...&r_{n-1}\\
r_1&...&r_l&r_{l+2}&...&r_n\\
\end{array}
\right)
$$

is the rational normal scroll (see p.106 [23]). The mixed states corresponding to them are as follows.\\

{\bf Example 5 (entangled mixed state from rational normal scroll).} Let $H=H_A^{n+1} \otimes H_B^2$ and $|\phi_1 
\rangle=\frac{1}{\sqrt{2}}(|01>+|12>)$,..,$|\phi_i\rangle=\frac{1}{\sqrt{2}}(|(i-1)1>+|i2>)$,...,$|\phi_l \rangle 
=\frac{1}{\sqrt{2}}(|(l-1)1>+|l2>)$,$|\phi_{l+1} \rangle=\frac{1}{\sqrt{2}}(|(l+1)1>+|(l+2)2>)$, ...,$|\phi_j 
\rangle=\frac{1}{\sqrt{2}}(|j1>+|(j+1)2>)$,...,$|\phi_{n-1} \rangle=\frac{1}{\sqrt{2}}(|(n-1)1>+|n2>)$.   We consider 
the mixed state $\rho_l=\frac{1}{n-1}(P_{|\phi_1 \rangle}+...+P_{|\phi_{n-1}\rangle})$  of rank $n-1$ on $H$. It is 
clear from section 2 $V_A^1(\rho)=X_{l,n-l-1} \subset CP^n$. From the well-known fact in algebraic geometry (see 
pp.92-93 of [23]) we have the following result.\\

{\bf Proposition 3.} {\em The mixed states $\rho_l, l=1,...,[\frac{n-1}{2}]$ are entangled and $\rho_l$ and 
$\rho_{l'}$ for $l\neq l'$ are not equivalent under local unitary transformations.}\\

We need to recall a well-known result in the theory of determinantal varieties (see Proposition in p.67 of [4]). Let 
$M(m,n)=\{(x_{ij}): 1\leq i \leq m, 1 \leq j \leq n\}$ (isomorphic to $CP^{mn-1}$) be the projective space of all $ m 
\times n$ matrices. For a integer $0 \leq k \leq min\{m,n\}$, $M(m,n)_k$ is defined as the locus $\{A=(x_{ij}) \in 
M(m,n): rank(A) \leq k\}$. $M(m,n)_k$ is called generic determinantal varieties.\\

{\bf Proposition 4.} {\em $M(m,n)_k$ is an irreducible algebriac subvariety of $M$ of codimension $(m-k)(n-k)$.}\\

Suppose $m \leq n$,  we now construct a mixed state $\rho$ with $V_A^{m-1}(\rho)=M(m,n)_{(m-1)}$.\\

{\bf Example 6 (generic entangled mixed state).} Let $H=H_A^{mn} \otimes H_B^m$,where $ m \leq n$,  and $A_{ij}$, 
$i=1,...,m,j=1,...n$ be $m \times n$ matrix with only nonzero entry at $ij$ position equal to 1. Let $A$ be a blocked 
$mn \times 1$ matrix with $ij$ block $A_{ij}$. Here the $k$-th row of $A_{ij}$ in $A$ corresponds to the vector 
$|(ij)k \rangle$ in the standard basis of $H$. Hence $A$ is a size $m^2n \times n$ matrix. Let 
$\rho=\frac{1}{D}AA^{\dagger}$ be a mixed state on $H$ (Here $D$ is a normalizing constant). It is a rank $n$ mixed 
state.\\

It is easy to compute $\Sigma_{ij}r_{ij} A_{ij}=(r_{ij})_{1\leq i \leq m, 1\leq j \leq n}$ (up to a constant). Thus we 
have $V_A^{m-1}(\rho)=M(m,n)_{(m-1)}$. From Proposition 4 and Theorem 3, $\rho$ is an entangled mixed state.\\

{\bf Example 7.} Let $H=H_A^m \otimes H_B^n \otimes H_C^m$, $|\phi_l \rangle=\frac{1}{m}\Sigma_{i=1}^m |ili \rangle$ 
for $l=1,...,n$ and $\rho=\frac{1}{n}(P_{|\phi_1 \rangle}+...+P_{|\phi_n \rangle})$ be a rank n mixed state on $H$. It 
is clear that under the cut B:AC, $\rho$ is separable. However, under the cut AB:C, we can check that $\rho$ is just 
the mixed state in Example 6 and thus entangled. Similarly under the cut A:BC, $\rho$ is also the mixed state in 
Example 6 and thus entangled. Hence this is a mixed state on tripartite quantum system with the property that it is 
separable under B:AC cut and entangled under AB:C and A:BC cuts.\\

{\bf Example 8 (entangled mixed states from Eisenbud Theorem [20]).} Let $H=H_A^h \otimes H_B^{m}$, where $h \geq 
nm-m+1$ and $n \geq m$ are positive integers, $A_i$, $i=1,...,h$ be $m \times n$ matrix with the property: the space 
$M$ of linear forms (of $r_1,...,r_h$) span by the entries of $T(r_1,...,r_h)=\Sigma_i r_i A_i$ is of dimension $h$. 
Let  $A$ be the $hm \times n$ matrix with $i$-th block $A_i$. Here the $k$-th row of $A_i$ in $A$ corresponds to the 
vector $|ik \rangle$ in the standard basis of $H$. Let $\rho=\frac{1}{D}AA^{\dagger}$  ( $D$ is a normalizing 
constant) be a (rank n) mixed state on $H$. From the proof of Theorem 2 we have $V_A^{m-1}(\rho)=\{(r_1,...,r_h): 
rank(T(r_1,...,r_h)) \leq m-1\}$\\

{\bf Theorem 6.} {\em $\rho$ is an entangled mixed state.}\\

{\bf Proof.} It is clear that $M(m,n)$ is 1-generic (see [20], p.548) We can see that the space $M$ has codimension 
(in $M(m,n)$) smaller or equal to $m-1$ (here $v=n,w=m,k=m-1$ as refered to Theorem 2.1 in p.552 of [20]). From 
definition it is clear that $V_A(\rho)$ is just $M_{m-1}$ , which is reduced and irreducible and of codimension $m-1$ 
in $M(m,n)$ from Theorem 2.1 of [20]. The conclusion is proved.\\

Eisenbud Theorem (Theorem 2.1 in [20] and thus Theorem 6  here) gives us a general method to construct many entangled 
states of low ranks, since the condition about $M$ is not a very strong restriction.\\ 

{\bf Example 9 (Bennett-DiVincenzo-Mor-Shor-Smolin-Terhal 's  mixed state from UPB [7] )} Let $H=H_A^2 \otimes H_B^2 
\otimes H_C^2$, $|\phi_{+} \rangle=\frac{1}{\sqrt{2}}(|1 \rangle+|2 \rangle), |\phi_{-} \rangle=\frac{1}{\sqrt{2}}(|1 
\rangle-|2 \rangle)$. Consider the linear subspace $T$ spaned by the following 4 vectors $|1 \rangle \otimes |2 
\rangle \otimes |\phi_{+} \rangle$, $|2\rangle \otimes |\phi_{+}\rangle \otimes |1 \rangle$, $|\phi_{+}\rangle \otimes 
|1 \rangle\otimes |2\rangle$, $|\phi_{-}\rangle \otimes |\phi_{-}\rangle \otimes |\phi_{-}\rangle$. Now $P$ is the 
projection to the complementary space $T^{\perp}$ of $T$ and $\rho=\frac{1}{D} P$ is a rank 4 PPT mixed state on $H$. 
It is proved in [7] that $\rho$ is entangled under the cut A:B:C (thus bound entanglement), however, it is separable 
under the cuts A:BC,B:AC,C:AB. Now we can compute its invariants $V_{AB}^1(\rho)$ and $V_{A:B}^1(\rho)$. It is easy to 
see from Theorem 3 that $V_{AB}^1(\rho)$ should be linear, however we can see that $V_{A:B}^1(\rho)$ is also linear 
from our computation below, though it is entangled under the cut A:B:C.\\

It is easy to check that the following 4 vectors $|010 \rangle-|011 \rangle$, $|100 \rangle-|110 \rangle$, $|001 
\rangle -|101 \rangle$, $|000 \rangle-|111 \rangle$ are the base of $T^{\perp}$. Thus the matrix $A$ is of the 
following form (with rows correspond to $|000 \rangle,|001 \rangle, 
|010\rangle,|011\rangle,|100\rangle,|101\rangle,|110\rangle,|111\rangle$)\\

$$
\left(
\begin{array}{cccccccc}
0&0&0&1\\
0&0&1&0\\
1&0&0&0\\
-1&0&0&0\\
0&1&0&0\\
0&0&-1&0\\
0&-1&0&0\\
0&0&0&-1
\end{array}
\right)
(15)
$$

and $\Sigma_{ij} r_{ij} A_{ij}$ is of the following form\\

$$
\left(
\begin{array}{cccccc}
r_{01}&r_{10}-r_{11}&0&r_{00}\\
-r_{01}&0&r_{00}-r_{10}&-r_{11}
\end{array}
\right)
(16)
$$

Thus $V_{AB}^1(\rho)$ is the union of the following 3 points $(1:1:0:0),(0:1:0:1),(0:0:1:0)$ in $CP^3$ and 
$V_{A:B}^1(\rho)$ is union of $CP^1 \times (1:0),(0:1) \times CP^1$ and $(1:0) \times (0:1)$ in $CP^1 \times CP^1$.\\

\section*{5. Generalized Smolin state}

J.A.Smolin [40] introduced a rank 4 mixed state $\rho=\frac{1}{4}(P_{|v_1\rangle}+P_{|v_2 \rangle}+P_{|v_3 
\rangle}+P_{|v_4 \rangle})$ on 4-party quantum system $H_A^2 \otimes H_B^2 \otimes H_C^3 \otimes H_D^2$, where,\\

$$
\begin{array}{ccccccccccccccccc}
|v_1\rangle = \frac{1}{2}(|0000 \rangle +|0011 \rangle + |1100 \rangle +|1111 \rangle)\\
|v_2\rangle = \frac{1}{2}(|0000 \rangle -|0011 \rangle - |1100 \rangle +|1111 \rangle)\\
|v_3\rangle = \frac{1}{2}(|0101 \rangle +|0110 \rangle + |1001 \rangle +|1010 \rangle)\\
|v_2\rangle = \frac{1}{2}(|0101 \rangle -|0110 \rangle - |1001 \rangle +|1010 \rangle)
\end{array}
$$

This mixed state $\rho$ is a bound entangled state when 4 parties A,B,C,D are isolated.\\

The following example, which is a continuous family (depending on 4 parameters) of  mixed state in the 4-party quantum 
system $H_A^2 \otimes H_B^2 \otimes H_C^2 \otimes H_D^2$ separable for any $2:2$ cut but entangled for any $1:3$ cut 
(thus bound entangled mixed state when A,B,C,D are isolated),  can be thought as a generalization of Smolin's mixed 
state in [40]. We prove that the generic members in this family of mixed states are not equivalent under local unitary 
transformations (Theorem 7 below).\\

{\bf Example 10.} Let $H=H_{A}^2 \otimes H_{B}^2 \otimes H_{C}^2 \otimes H_{D}^2$ and $h_1,h_2,h_3,h_4$ (understood as 
row vectors) be 4 mutually orthogonal unit vectors in $C^4$ and $a_1,a_2,a_3,a_4$ be 4 parameters. Consider the $16 
\times 4$ matrix $T$ with 16 rows as\\
 $T=(a_1h_1^{\tau},0,0,a_2 h_2^{\tau},0, a_3 h_3^{\tau},a_4 h_4^{\tau},0,0, a_5 h_4^{\tau}, a_6 h_3^{\tau},0, a_7 
h_2^{\tau},0,0,a_8 h_1^{\tau})^{\tau}$. Let $|\phi'_1 \rangle$, $|\phi'_2 \rangle$, $|\phi'_3 \rangle$, $|\phi'_4 
\rangle$ be 4 vectors in $H$ whose expansions with the basis $|0000\rangle,|0001\rangle$, 
$|0010\rangle,|0011\rangle,|0100\rangle$, $|0101\rangle,|0110\rangle,|0111\rangle,|1000\rangle$, 
$|1001\rangle,|1010\rangle,|1011\rangle$, $|1100\rangle$,
 $|1101\rangle,|1110\rangle,|1111\rangle $ are exactly the 4 columns of the matrix $T$ and  
$|\phi_1\rangle,|\phi_2\rangle,|\phi_3\rangle$, $|\phi_4\rangle$ be the normalized unit vectors of $|\phi'_1 \rangle, 
|\phi'_2\rangle, |\phi'_3\rangle$, $|\phi'_4\rangle$. Let $\rho=\frac{1}{4}(P_{|\phi_1 \rangle} +P_{|\phi_2\rangle} 
+P_{|\phi_3 \rangle} +P_{|\phi_4 \rangle})$.\\

It is easy to check that when $h_1=(1,1,0,0),h_2=(1,-1,0,0), h_3=(0,0,1,1)$, $h_4=(0,0,1,-1)$ and $a_1=a_2=a_3=a_4=1$. 
It is just the Smolin's mixed state in [40].\\

Now we prove that $\rho$ is invariant under the partial transposes of the cuts AB:CD,AC:BD,AD:BC.\\

Let  the ``representation'' matrix $T=(b_{ijkl}^t)_{i=0,1,j=0,1,k=0,1,l=0,1,t=1,2,3,4}$ be the matrix with columns 
corresponding the expansions of $\phi_1,\phi_2,\phi_3,\phi_4$.Then we can consider that $T=(T_1,T_2,T_3,T_4)^{\tau}$ 
as a blocked matrix of size $4 \times 1$ with each block $T_{ij}=(b_{kl }^t)_{k=0,1,l=0,1,=1,2,3,4}$ a $4 \times 4$ 
matrix, where $ij=00,01,10,11$. Because $h_1,h_2,h_3,h_4$ are mutually orthogonal unit vectors we can easily check 
that $T_{ij} T_{i'j'}^{\dagger}=T_{i'j'} T_{ij}^{\dagger}$ Thus it is invariant when the partial transpose of the cut 
AB:CD is applied.\\

With the same methods we can check that $\rho$ is invariant when the partial transposes of the cuts AC:BD, AD:BC  are 
applied. Hence $\rho$ is PPT under the cuts AB:CD, AC:BD,AD:BC. Thus from a result in [30] we know $\rho$ is separable 
under these cuts AB:CD, AC:BD,AD:BC.\\

Now we want to prove $\rho$ is entangled under the cut A:BCD by computing $V_{BCD}^1(\rho)$. From the previous 
arguments , we can check that $V_{BCD}^1(\rho)$ is the locus of the condition: $a_1 h_1 r_{000} + a_2 h_2 r_{011} +a_3 
h_3 r_{101} +a_4 h_4 r_{110}$ and $a_7 h_1 r_{100} + a_8 h_2 r_{111} +a_5 h_3 r_{001} +a_6 h_4 r_{010}$ are linear 
dependent. This is equivalent to the condition that the matrix (12) is of  rank 1.\\

$$
\left(
\begin{array}{cccccc}
a_7 r_{100} & a_8 r_{111} & a_5 r_{001} & a_6 r_{010}\\
a_1 r_{000} & a_2 r_{011} & a_3 r_{101} & a_4 r_{110}
\end{array}
\right)
$$

From [23] pp. 25-26 we can check that $V_{BCD}^1(\rho)$ is exactly the famous Segre variety in algebraic geometry. It 
is irreducible and thus cannot be linear. From Theorem 3, $\rho$ is entangled under the cut A:BCD. Similarly we can 
prove that $\rho$ is entangled under the cuts B:ACD, C:ABD, D:ABC.\\

Now we compute $V_{A:B}^3(\rho)$. From the previous arguments  and Definition 2 , it is just the locus of the 
condition that the vectors $h_1(a_1 r_0^1 r_0^2 +a_7 r_1^1 r_1^2)$, $h_3 (a_3 r_0^1 r_1^2 +a_5 r_1^1 r_0^2)$, $h_4(a_4 
r_0^1 r_1^2 +a_6  r_1^1 r_0^2)$, $h_2 (a_2 r_0^1 r_0^2 +a_8 r_1^1 r_1^2)$ are linear dependent. Since 
$h_1,h_2,h_3,h_4$ are mutually orthogonal unit vectors,we have \\

$$
\begin{array}{cccccccccc}
V_{A:B}^3(\rho)=\{(r_0^1,r_1^1,r_0^2,r_1^2) \in CP^1 \times CP^1:\\
(a_1 r_0^1 r_0^2 +a_7 r_1^1 r_1^2)(a_3 r_0^1 r_1^2 +a_5 r_1^1 r_0^2)(a_4 r_0^1 r_1^2 +a_6  r_1^1 r_0^2)(a_2 r_0^1 
r_0^2 +a_8 r_1^1 r_1^2)=0\}
\end{array}
$$

From Theorem 3 we know that $\rho$ is entangled for the cut A:B:CD, A:C:BD and A:D:BC for generic parameters, since 
$a_1r_0^1r_0^2+a_7r_1^1r_1^2$,etc., cannot be factorized to 2 linear forms for generic $a_1$ and $a_7$.  This provides 
another proof that the mixed state is entangled if the 4 parties are isolated.\\

Let $\lambda_1=-a_1/a_7,\lambda_2=-a_3/a_5, \lambda_3=-a_4/a_6, \lambda_4=-a_2/a_8$ and consider the family of the 
mixed states $\{\rho_{\lambda_{1,2,3,4}}\}$, we want to prove the following statement.\\

{\bf Theorem 7.} {\em The generic memebers in this continuous family of mixed states are inequivalent under the local 
unitary transformations on $H=H_{A}^2 \otimes H_{B}^2 \otimes H_{C}^2 \otimes H_{D}^2$.}\\

{\bf Proof.} From the above computation, $V_{A:B}^3(\rho_{\lambda_{1,2,3,4}})$ is the union of the following 4 
algbraic varieties in $CP^1 \times CP^1$.\\

$$
\begin{array}{ccccccccccc}
V_1=\{(r_0^1,r_1^1,r_0^2,r_1^2) \in CP^1 \times CP^1:r_0^1 r_0^2 - \lambda_1 r_1^1 r_1^2=0\}\\
V_2=\{(r_0^1,r_1^1,r_0^2,r_1^2) \in CP^1 \times CP^1:r_0^1 r_1^2 - \lambda_2 r_1^1 r_0^2=0\}\\
V_3=\{(r_0^1,r_1^1,r_0^2,r_1^2) \in CP^1 \times CP^1:r_0^1 r_1^2 - \lambda_3 r_1^1 r_0^2=0\}\\
V_4=\{(r_0^1,r_1^1,r_0^2,r_1^2) \in CP^1 \times CP^1:r_0^1 r_0^2 - \lambda_4 r_1^1 r_1^2=0\}
\end{array}
$$

 From Theorem 1', if $\rho_{\lambda_{1,2,3,4}}$ and $\rho_{\lambda'_{1,2,3,4}}$ are equivalent by a local operation, 
there must exist 2 fractional linear transformations $T_1, T_2$ of $CP^1$ such that $T=T_1 \times T_2$ (acting on  
$CP^1 \times CP^1$) transforms the 4 varieties $V_1,V_2,V_3,V_4$  of $\rho_{\lambda_{1,2,3,4}}$ to the  4 varieties 
$V'_1,V'_2,V'_3,V'_4$  of $\rho_{\lambda'_{1,2,3,4}}$,i.e., $T(V_i)=V'_j$.\\

Introduce the inhomogeneous coordinates $x_1=r_0^1/r_1^1,x_2=r_0^2/r_1^2$. Let $T_1(x_1)=(ax_1+b)/(cx_1+d)$.  Suppose 
$T(V_i)=V'_i, i=1,2,3,4$. Then we have $ab \lambda_1= cd \lambda'_1 \lambda'_2$ and $ab \lambda_4=cd \lambda'_3 
\lambda'_4$. Hence $\lambda_1 \lambda'_3 \lambda'_4 =\lambda'_1 \lambda'_2 \lambda_4$. This means that there are some 
algebraic relations of parameters if the $T$ exists. Similarly we can get the same conclusion for the other 
possibilities $T(V_i)=V'_j$. This implies that there are some algebraic relations of parameters $\lambda_{1,2,3,4}$ 
and $\lambda'_{1,2,3,4}$ if $\rho_{\lambda_{1,2,3,4}}$ and $\rho_{\lambda'_{1,2,3,4}}$ are  equivalent  by a local 
unitary transformations. Hence our conclusion follows immediately.\\

\section*{6. Non empty theorem}

In this section, we prove that the algebraic set invariants introduced in section 2 are not empty for low rank mixed 
states.\\

{\bf Theorem 8.} {\em Let $H=H_A^m \otimes H_B^n$ be a bipartite quantum system and $\rho$ is a rank $r$ mixed state 
on $H$ with $r \leq m+n-2$. Then $V_A^{n-1}(\rho)$ and $V_B^{m-1}(\rho)$ are not empty.}\\

{\bf Proof.} We take ``the standard representation''  $\rho=\Sigma_{i=1}^r p_i P_{v_i}$, where $p_1,...,p_r$, 
,$v_1,...,v_r$ are eigenvalues and eigenvectors of $\rho$ and $r=rank(\rho)$. Recall the proof of Theorem 2, 
$V_A^{n-1}(\rho)$ is the locus of the condition that $ \Sigma_i r_i A_i$ has its rank smaller than $n$. Thus from 
Proposition 4 above we know that the codimension of $V_A^{n-1}(\rho)$ in $CP^{m-1}$ is smaller or equal to 
$(n-(n-1))(r-(n-1))=r-n+1$. Hence $dim(V_A(\rho)) \geq m-1-r+n-1 \geq 0$ and $V_A^{n-1}(\rho)$ is not empty. The 
conclusion for $V_B^{m-1}(\rho)$ can be proved similarly.\\

\section*{7. A relation of determinants}

As indicated in section 2, we can have the following statement from Lemma 2.\\

{\bf Theorem 9.} {\em Let $H=H_A^m  \otimes H_B^n$ be a bipartite quantum system and $\rho=\Sigma_{l=1}^t p_l 
P_{v_l}=\Sigma_{l=1}^s q_l P_{v'_l}$ be a mixed state with two ``representations'' as convex combinations of 
projections with $p_1,...,p_t, q_1,...,q_s > 0$. Let $A$ (resp. $A'$ ) be the $mn \times t$ (resp. $mn \times s$) 
matrix of vectors $v_1,...,v_t$ (resp. $v'_1,...,v'_s$) in the standard basis 
$|11\rangle,...,|1n\rangle,...,|m1\rangle,...,|mn\rangle$ as in Lemma 1. We represent $A$ (resp.$A'$) as $m\times 1$ 
blocked matrix with blocks $A_1,...,A_m$ (resp.$A'_1,...,A'_m$). Then the determinantal varieties defined by the 
conditions that the ranks of $R=\Sigma_i r_i A_i$ and $R'=\Sigma_i r_i A'_i$ are smaller than $k$,$k=0,1,...,n-1$,  
are the same.}\\

Actually we can get more information about the determinants of $n \times n$ submatrices of $\Sigma_i r_i A_i$ and 
$\Sigma_i r_i A'_i$ from the proof of Theorem 2 and 3. This relation seems to be helpful to extract information of  
$\rho$'s one unknown ``representation'' from its another known ``representation'', as in the proof of Theorem 3.\\

{\bf Theorem 9'.} {\em Let $H, \rho,p_1,...p_t,q_1,...,q_s, A,A',R,R'$ be as above and $R_{i_1,...i_n}$ (resp. 
$R'_{i'_1,...,i'_n}$)  be the $n \times n$ submatrix of $R$ (resp. $R'$) consisting of $i_1<...<i_n$ 
(resp.$i'_1<...<i'_n$)-th columns, where $i_1,...,i_n \in \{1,...t\}$ and $i'_1,...,i'_n \in \{1,...,s\}$ are distinct 
indices. Then we have\\

$$
\begin{array}{ccccccccccccccccccc}
\Sigma_{i_1<...<i_n} p_{i_1}...p_{i_n} 
|R_{i_1,...,i_n}|^2=\Sigma_{i'_1<...<i'_n}q_{i'_1}...q_{i'_n}|R'_{i'_1,...,i'_n}|^2
\end{array}
(17)
$$
.}

The above result follows from the following Lemma immediately, Since both sides of the equality are just 
$det(\Sigma_{ij} r_i r_j^{\dagger} \rho_{ij})$.\\

{\bf Lemma 3 (Binet-Cauchy formula).} {\em Let $B$ be a $n\times t$ matrix with $t>n$ and $B_{i_1,...,i_n}$ be the $n 
\times n$ submatrix of $B$ consisting of $i_1<...<i_n$-th columns. Then $det(BB^{\dagger})=\Sigma_{i_1<...<i_n} |det 
B_{i_1,..,i_n}|^2$.}\\

It is clear that Theorem 2 and 3 follow from Theorem 8' here immediately.\\

The following result was  previously known in [29],[30].\\

{\bf Proposition 5.} {\em Let $H=H_A^n \otimes H_B^n$ , $\rho =\frac{1}{n}\Sigma_{i=1}^n p_i P_{a_i \otimes b_i}$ 
where $a_1,...,a_n$ (resp. $b_1,...,b_n$) are linearly independent unit vectors in $H_A^n$ (resp. $H_B^n$). Suppose 
that $\rho=\frac{1}{t}\Sigma_{i=1}^t q_i P_{c_i \otimes d_i}$ is another representation of $\rho$ as a convex 
combination with $q_i$'s positive, then actually we have $t=n$ and $\{ a_1 \otimes b_1,...,a_n \otimes b_n \}=\{c_1 
\otimes d_1,...,c_n \otimes d_n \}$.}\\

{\bf Proof.} We apply Theorem 9' to the 2 ``representations''  here. First of all,  we know that $\Sigma_{ij} r_i 
r_j^{\dagger} \rho_{ij}$ is (up to a nonzero constant)the square of the absolute value of a multiplication of linear 
forms $b_i(r_1,...,r_n)=\Sigma_j b_1^j r_j$,  where $b_i=\Sigma_j b_i^j |j>$ is the coordinate form of $b_i$ for 
$i=1,...,n$,  from one known ``representation''. Thus we know from Theorem 8' that there are at least $n$ vectors in 
$\{d_1,...,d_t\}$ , without loss of generality,  suppose they are $d_1,...,d_n$,  are just $b_1,...,b_n$. Using 
Theorem 9' for the second factor and consider the 1st,...,n-th columns of $R'$ , this implies that the multiplication 
of the linear forms $c_1(r_1,...r_n),...,c_n(r_1,...,r_n)$ are just the multiplication of the linear forms 
$a_1(r_1,...,r_n),...,a_n(r_1,...,r_n)$. Hence we know that the set  $\{c_{1},...,c_{n}\}$ are just the set 
$\{a_1,...,a_n\}$. \\

On the other hand it is easy to see that $a_i \otimes b_j$ with $i \neq j$ is not in the linear span of $a_1 \otimes 
b_1,...,a_n \otimes b_n$, since $a_1,..,a_n$ (resp. $b_1,...,b_n$)  are linear independent. Thus $c_i=a_i$ from 
Corollary 1.\\

Applying Theorem 9' to other columns of $R$ and $R'$ by a similar argument, we have  $c_j \in \{a_1,...,a_n\}$ and 
$d_j \in \{b_1,...,b_n\}$. Since $a_i \otimes b_j$ with $i \neq j$ cannot be in the image of $\rho$, $c_j \otimes d_j$ 
has to be the form $a_{i_j} \otimes b_{i_j}$. The conclusion is proved.\\

{\bf Remark 2.} If we compute $V_A^{n-1}(\rho)$ from the representation of $\rho$ 's standard form, i.e., linear sum 
of projections to its eigenvectors, it can be seen that our invariants defined in section 2 are independent of 
eigenvalues ($p_1,...,p_t$ in section 2). However the information of $p_1,...,p_t$ or eigenvalues is certainly 
reflected in Theorem 9' here. Thus Theorem 9' might be more useful in determining whether a given mixed state is 
entangled or not, provided that  we know how to extract sufficient information from Theorem 9'.\\

{\bf Remark 3.} As showed in Example 1, our invariants might be empty set for high rank mixed states, however it seems 
that Theorem 9' is still useful in determining whether a given high rank mixed state is entangled or not in this case, 
provided that we know how to extract information from Theorem 9'.\\

\section*{8. Simulation of Hamiltonians}

Historically, the idea of simulating Hamiltonian (self-adjoint operators on  the Hilbert space corresponding to the 
quantum system , see [39]) time evolutions was the first motivation for quantum computation because of the famous 
paper of Feynman. Recently the ability of nonlocal Hamiltonians to simulate one another is a popular topic , which has 
applications in quantum control theory, quantum compuation and the task of generating enatnglement . For the general 
treatment of this topic and the references, we refer to [9].\\

We say, for two bipartite Hamiltonians $H$ 
and $H'$ on $H_A^m \otimes H_B^n$, $H'$ can be efficiently 
simulated by $H$ with local unitary operations, write as 
$H' \prec_{LU} H$, if $H'$ can be written as a convex 
combination of conjugates of $H$ by local unitary operations, 
$H'=p_1(U_1 \otimes V_1)H(U_1 \otimes V_1)^{\dagger}+...
+p_s (U_s \otimes V_s) H(U_s \otimes V_s)^{\dagger}$, 
where $p_1,...,p_s$ are positive real numbers such that 
$p_1+...+p_s=1$ , $U_1,...,U_s$ and $V_1,...,V_s$ are 
unitary operations on $H_A^m$ and $H_B^n$ respectively.
 Here we use $\dagger$ for the adjoint. This is equivalent to the notion
"infinitesimal simulation" in [9]. In [9]  it is shown that "local terms" like $I \otimes K_B$ and $K_A \otimes I$ are 
irrelevant to the simulation problem up to the second order, thus they considered the simulation problem for 
Hamiltonians 
without local terms' effect. Our definition here is more restricted without neglecting the local terms.\\

We can have the following necessary conditions about the simulation of semi-positive Hamiltonians based on the 
algebraic set invariants introduced in section 2.\\

{\bf Theorem 10.} {\em Let $H$ and $H'$ be two semi-positive Hamiltonians in the bipartite quantum system $H_A^m 
\otimes H_B^n$. Suppose  $H' \prec_{LU} H$, that is , $H'$ can be simulated by $H$ efficiently by using local unitary 
transformations. Then there are projective isomorphisms $U_1$ of $CP^{m-1}$ and $V_1$ of $CP^{n-1}$ such that 
$U_1(V_A^k(H')) \subset V_A^k(H)$ for $k=0,...,n-1$ and $V_1(V_B^k(H')) \subset V_B^k(H)$ for $k=0,...,m-1$.}\\

The following observation about the computation of $V_A^k(\rho)$ is the the key point of the proof of Theorem 10 and 
Corollary 3. From Corollary 1 if $\rho=\Sigma_i^t p_i P_{v_i}$ with $p_i$'s positive real numbers, the range of $\rho$ 
is the linear span of vectors $v_1,...,v_t$. We take some vectors in the set $\{v_1,...,v_t\}$, say they are 
$v_1,...,v_s$. Let $B$ be the $mn \times s$ matrix with columns corresponding to the $s$ vectors $v_1,...,v_s$'s 
coordinates in the standard basis of $H_A^m \otimes H_B^n$. We consider $B$ as $m \times 1$ blocked matrix with blocks 
$B_1,...,B_m$ $n \times s$ matrix as in section 2. It is clear that $V_A^k(\rho)$ is an algebraic subset of the zero 
locus of the determinants of all $(k+1) \times (k+1)$ sub-matrices of $\Sigma_i^mr_iB_i$, since $\Sigma_i^m r_iB_i$ is 
a sub-matrix of $\Sigma_i^mr_i A_i$. On the other hand if $v_1,...,v_s$ are linear independent and 
$s=dim(range(\rho))$, $V_A^k(\rho))$ is just the zero locus of the determinants of all $(k+1) \times (k+1)$ 
sub-matrices of $\Sigma_i^m r_i B_i$, since any column in $\Sigma_i r_i A_i$ is a linear combination of columns in 
$\Sigma_i r_i B_i$ ($rank(\Sigma_i r_i A_i) \leq k$ is equivalent to $rank(\Sigma_i r_i B_i) \leq k$).\\

{\bf Proof of Theorem 10.} Suppose $H' \prec_{LU} H$, then there exist positive numbers $p_1,...,p_s$ and local 
unitary transformations $U_1 \otimes V_1,...,U_s \otimes V_s$, such that, $\Sigma_i p_i U_i \otimes V_i H (U_i \otimes 
V_i)^{\dagger}=H'$. Let $H=\Sigma_i^s q_i P_{\psi_i}$, where $s=dim(range(H))$ , $q_1,...,q_s$ are eigenvalues of $H$ 
and $\psi_1,...,\psi_s$ are eigenvectors of $H$. Then it is clear that $(U_i \otimes V_i)H(U_i \otimes 
V_i)^{\dagger}=\Sigma_j^s q_j P_{(U_i \otimes V_i)\psi_j}$ and thus $H'=\Sigma_{i,j} p_i q_j P_{(U_i \otimes 
V_i)\psi_j}$. This is a representation of $H'$ as a convex combination of projections. From our above observation 
$V_A^k(H')$  is an algebraic subset in $V_A^k((U_1 \times V_1)H)$ (which can be computed from vectors $(U_1 \otimes 
V_1)\psi_1,...,(U_1 \otimes V_1)\psi_s$). Thus the conclusion follows from Theorem 1.\\

{\bf Corollary 3} (see [13]). {\em Let $H$ and $H'$ be two semi-positive Hamiltonians in the bipartite quantum system 
$H_A^m \otimes H_B^n$ of the same rank, i.e. $dim(range(H))=dim(range(H'))$. Suppose  $H' \prec_{LU} H$, that is , 
$H'$ can be simulated by $H$ efficiently by using local unitary transformations. Then $V_A^k(H')=V_A^k(H)$ for 
$k=0,...,n-1$ and $V_B^k(H')=V_B^k(H)$ for $k=0,...,m-1$. Here the equality of the algebraic sets means they are 
isomorphic via  projective linear transformations of complex projective spaces.}\\

{\bf Proof.} Suppose $H' \prec_{LU} H$, then there exist positive numbers $p_1,...,p_s$ and local unitary 
transformations $U_1 \otimes V_1,...,U_s \otimes V_s$, such that, $\Sigma_i p_i U_i \otimes V_i H (U_i \otimes 
V_i)^{\dagger}=H'$. Let $H=\Sigma_i^s q_i P_{\psi_i}$, where $s=dim(range(H))$ , $q_1,...,q_s$ are eigenvalues of $H$ 
and $\psi_1,...,\psi_s$ are eigenvectors of $H$. Then it is clear that $(U_i \otimes V_i)H(U_i \otimes 
V_i)^{\dagger}=\Sigma_j^s q_j P_{(U_i \otimes V_i)\psi_j}$ and thus $H'=\Sigma_{i,j} p_i q_j P_{(U_i \otimes 
V_i)\psi_j}$. This is a representation of $H'$ as a convex combination of projections. From our above observation 
$V_A^k(H')$ can be computed from vectors $(U_1 \otimes V_1)\psi_1,...,(U_1 \otimes V_1)\psi_s$ , since they are linear 
independent and $s=dim(range(H'))$. Hence $V_A^k(H')=V_A^k((U_1 \otimes V_1)H)$ from the definition. Thus the 
conclusion follows from Theorem 1.\\

Let $S$ be the swap operator on the bipartite system $H_A^n \otimes H_B^n$ defined 
by $S|ij\rangle=|ji\rangle$. For any Hamiltonian $H$, $S(H)=SHS^{\dagger}$ corresponds to the Hamiltonian evolution of
 $H$ with A and B interchanged. It is very interesting to consider the problem if $H$ can be simulated by $S(H)$ 
efficiently .
 This led to some important consequences in the discussion VII of [9]. For example it was shown there
 are examples that $H$ and $S(H)$ cannot be simulated efficiently with one another in higher dimensions. Thus in 
higher 
dimensions nonlocal degrees of freedom of Hamiltonians cannot be characterized by quantities that are symmetric with
 respect to A and B, such as eigenvalues. This conclusion  is also obtained from the our example and Corollary 5 in 
the next section. From Corollary 3 we have the following necessary condition about $H \prec_{LU} S(H)$.\\

{\bf Corollary 4.} {\em Let $H$ be a semi-positive Hamiltonian on $H_A^n \otimes H_B^n$. 
Suppose $H \prec_{LU} S(H)$. Then $V_A^k(H)=V_B^k(H)$ for $k=0,...,n-1$.}\\

The following is a Hamiltonian $H$  on $3 \times 3$ system for which $H$ cannot be 
simulated efficiently by $S(H)$.\\

{\bf Example 11.} 
$H=P_{|\phi_1 \rangle}+ P_{|\phi_2 \rangle }+P_{|\phi_3 \rangle}$, where
\begin{equation}
\begin{array}{cccccc}
|\phi_1 \rangle=\frac{1}{\sqrt{3}}(|11\rangle +|21\rangle +|32\rangle)\\
|\phi_2 \rangle=\frac{1}{\sqrt{1+|v|^2}}(|12\rangle+v|22\rangle)\\
|\phi_3 \rangle=\frac{1}{\sqrt{1+|\lambda|^2}}(|13\rangle +\lambda|23\rangle)\\
\end{array}
\label{(2)}
\end{equation}

Then it is easy to compute that $V_A^2(H)$ is the sum of 3 lines in $CP^2$ 
defined by $r_1+r_2=0$,$r_1+vr_2=0$ and $r_1+\lambda r_2=0$ for $v \neq \lambda$ and both $v ,\lambda$ are not 1, 
and $V_B^2(H)$ is the sum of 2 lines in $CP^2$ defined by $r_2=0$ and $r_3=0$. Thus we cannot have $H \prec_{LU} 
S(H)$.

\section*{9. A continuous family of states and Hamiltonians related to elliptic curves}

From physical point of view, it is very interesting to have isospectral (i.e., eigenvalues of $\rho, 
tr_A(\rho),tr_B(\rho)$ are the same) mixed states , but they are not equivalent under local unitary transformations. 
This phenomenon indicates that we cannot obetain a complete understanding of a bipartite quantum system by just 
studying the local and global properties of the spectra of the system. Some examples of such mixed states have been 
found by several authors(see Nielsen and Kempe [37] and references there). Here we give a continuous family of such 
mixed states.\\

Let $H=H_A^3 \otimes H_B^3$ and  $\rho_{\eta_1,\eta_2,\eta_3}=\frac{1}{3}(P_{|v_1 \rangle}+P_{|v_2 \rangle}+P_{|v_3 
\rangle})$ ($\eta_1,\eta_2,\eta_3$ are real parameters), a continuous family of mixed states on $H$, where \\

$$
\begin{array}{cccccccc}
|v_1 \rangle=\frac{1}{\sqrt{3}}(e^{i\eta_1}|11 \rangle+|22 \rangle+|33 \rangle)\\
|v_2 \rangle=\frac{1}{\sqrt{3}}(e^{i\eta_2}|12 \rangle+|23 \rangle+|31 \rangle)\\
|v_3 \rangle=\frac{1}{\sqrt{3}}(e^{i\eta_3}|13 \rangle+|21 \rangle+32 \rangle)
\end{array}
(19)
$$

It is easy to calculate that $\Sigma r_i A_i$ (up to a constant) is the following $3 \times 3$ matrix\\

$$
\left(
\begin{array}{cccccc}
e^{i\eta_1}r_1&r_3&r_2\\
r_2&e^{i\eta_2}r_1&r_3\\
r_3&r_2&e^{i\eta_3}r_1
\end{array}
\right)
$$

Thus $V_{A}^2(\rho_{\eta_1,\eta_2,\eta_3})$ is defined by 
$e^{i(\eta_1+\eta_2+\eta_3)}r_1^3+r_2^3+r_3^3-(e^{i\eta_1}+e^{i\eta_2}+e^{i\eta_3})r_1 r_2 r_3=0$ in $CP^{2}$. With 
$e^{i(\eta_1+\eta_2+\eta_3)/3}r_1=r_1'$ we have 
$r_1'^{3}+r_2^3+r_3^3-\frac{e^{i\eta_1}+e^{i\eta_2}+e^{i\eta_3}}{e^{i(\eta_1+\eta_2+\eta_3)/3}}r_1'r_2r_3=0$. This is 
a  family of elliptic curves.\\

It is easy to check that 3 nonzero eigenvalues of 
$\rho_{\eta_1,\eta_2,\eta_3},tr_A(\rho_{\eta_1,\eta_2,\eta_3}),tr_B(\rho_{\eta_1,\eta_2,\eta_3})$ are all the same 
value $\frac{1}{3}$ for different parameters. In this case  we have a family of isospectral (both global and local) 
mixed states $\rho_{\eta_1,\eta_2,\eta_3}$. Set 
$g(\eta_1,\eta_2,\eta_3)=\frac{e^{i\eta_1}+e^{i\eta_2}+e^{i\eta_3}}{e^{i(\eta_1+\eta_2+\eta_3)/3}}$.\\

{\bf Theorem 11.} {\em $\rho_{\eta_1,\eta_2,\eta_3}$ is entangled mixed state when  $(g(\eta_1,\eta_2,\eta_3))^3 \neq 
0,-216,27$. Moreover $\rho_{\eta_1,\eta_2,\eta_3}$ and $\rho_{\eta'_1,\eta'_2,\eta'_3}$ are not equivalent under local 
unitary transformations if $k(g(\eta_1,\eta_2,\eta_3)) \neq k(g(\eta'_1,\eta'_2,\eta'_3))$, where 
$k(x)=\frac{x^3(x^3+216)^3}{(-x^3+27)^3}$ is the moduli function of elliptic curves.}\\

{\bf Proof.} The conclusion follows from Theorem 3,1 and the well-known fact about elliptic curves (see [10]) \\

From Theorem 11 we give continuous many isospectral non-local-equivalent rank 3 mixed states in $H$.\\

In [36] Nielsen gave a beautiful necessary and sufficient condition for the bipartite pure state $|\psi \rangle$ can 
be transformed to the pure state $|\phi \rangle$ by local operations and classical communications (LOCC) based on the 
majorization between the eigenvalue vectors of the partial traces of $|\psi \rangle$ and $|\phi \rangle$. In [8] an 
example was given, from which we know that Nielsen's criterion cannot be generalized to multipartite case, {\bf 3EPR} 
and {\bf 2GHZ} are understood as pure states in a $4 \times 4 \times 4$ quantum system, they have the same eigenvalue 
vectors when traced over any subsystem. However it is proved that they are LOCC-incomparable in [8]\\

In the following example, a continuous family $|\{\phi\}_{\eta_1,\eta_2,\eta_3} \rangle$ of pure states in tripartite 
quantum system $H_{A_1}^3 \otimes H_{A_2}^3 \otimes H_{A_3}^3$ is given, the eigenvalue vectors of 
$tr_{A_i}(P_{|\phi_{\eta_1,\eta_2,\eta_3} \rangle}), tr_{A_iA_j}(P_{|\phi_{\eta_1,\eta_2,\eta_3} \rangle })$ are 
independent of parameters $\eta_1,\eta_2,\eta_3$. However the {\em generic}  pure states in this family are entangled 
and  LOCC-incomparable. This gives stronger evidence that it is  hopeless to characterize the entanglement properties 
of multipartite pure states by only using the eigenvalue spetra of their partial traces.\\

Let $H=H_{A_1}^3 \otimes H_{A_2}^3 \otimes H_{A_3}^3$ be a tripartite quantum system and $|\phi_{\eta_1,\eta_2,\eta_3} 
\rangle=\frac{1}{\sqrt{3}}(|v_1 \rangle \otimes |1 \rangle+|v_2 \rangle \otimes |2\rangle+|v_3\rangle \otimes 
|3\rangle)$, where $|v_1 \rangle, |v_2 \rangle,|v_3 \rangle$ are as in (19). This is a continuous family of pure 
states in $H$ parameterized by three real parameters. We can check that the eigenvalue vector of any partial trace of 
$P_{|\phi_{\eta_1,\eta_2,\eta_3} \rangle}$ is a constant vector. On the other hand it is clear that 
$tr_{A_3}(P_{|\phi_{\eta_1,\eta_2,\eta_3} \rangle})=\frac{1}{3}(P_{|v_1 \rangle}+P_{|v_2 \rangle}+P_{|v_3 \rangle})$ 
is a rank 3 mixed state in $H_{A_1}^3 \otimes H_{A_2}^3$. $|\phi_{\eta_1,\eta_2,\eta_3} \rangle$ and 
$|\phi_{\eta'_1,\eta'_2,\eta'_3} \rangle$ are not equivalent under local unitary transformations if 
$k(g(\eta_1,\eta_2,\eta_3)) \neq k(g(\eta'_1,\eta'_2,\eta'_3))$,   since their corresponding traces over $A_3$ are not 
equivalent under local unitary transformations of $H_{A_1}^3 \otimes H_{A_2}^3$ from Theorem 11. Hence the {\em 
generic} members of this family of pure states in tripartite quantum system $H$ are enatngled and  LOCC-incomparable 
from Theorem 1 in [8]. \\

We can also consider the following continuous family of semi-positive Hamiltonians depending on 3 real parameters, 
$H_{\eta_1,\eta_2,\eta_3}=P_{|v_1 \rangle} + P_{|v_2 \rangle}+P_{|v_3 \rangle}$, where $v_1,v_2,v_3$ are as in (19). 
As calculated above, $V_A^2(H_{\eta_1,\eta_2,\eta_3})$ is just the elliptic curve in $CP^2$ defined by $r_1^3 +r_2^3 
+r_3^3 -\frac{e^{i\eta_1}+e^{i\eta_2}+e^{i\eta_3}}{e^{i(\eta_1+\eta_2+\eta_3)/3}}r_1r_2r_3=0$. The elliptic curve 
$V_A^2(H_{\eta_1,\eta_2,\eta_3})$ is not isomorphic to the elliptic curve $V_A^2(H_{\eta'_1,\eta'_2,\eta'_3})$ if 
$k(g(\eta_1,\eta_2,\eta_3)) \neq k(g(\eta'_1,\eta'_2,\eta'_3))$. Thus we have the following Corollary of Theorem 9.\\

{\bf Corollary 5.} {\em $H_{\eta'_1,\eta'_2,\eta'_3}$ cannot be simulated by $H_{\eta_1,\eta_2,\eta_3}$ efficiently by 
using local unitary transformations,i.e.,we cannot have  $H_{\eta'_1,\eta'_2,\eta'_3} 
\prec_{LU}H_{\eta_1,\eta_2,\eta_3}$, if $k(g(\eta_1,\eta_2,\eta_3)) \neq k(g(\eta'_1,\eta'_2,\eta'_3))$, though the 3 
nonzero eigenvalues of  $H_{\eta_1,\eta_2,\eta_3}$, $H_{\eta'_1,\eta'_2,\eta'_3}$ and their partial traces are all 
1.}\\

\section*{10. Constructing entangled PPT mixed states}

As mentioned in Introduction, the first several entangled PPT mixed states were constructed in [25] based on P. 
Horodecki's range criterion of separable states, which asserts that a separable mixed state has to include 
sufficiently many {\em separable pure states} in its own range (see [25],[28]). This range criterion of separable 
mixed states  was also the base to construct PPT entangled mixed states in the context of unextendible product base 
(UPB) studied by C.H.Bennett, D.P.DiVincenzo, T.Mor, P.Shor, J.A.Smolin  and T.M.Terhal in [7] (We should mention that 
unextendible product base also have other physical significance {\em nonlocality without entanglement}, see [7],[28]). 
It is always interesting and important to have more methods to construct entangled PPT mixed states. In this section, 
we give  an example to show how our separability criterion Theorem 3 can be used to construct entangled mixed states 
which are invariant under partial transposition (thus PPT and bound entanglement) systematically.\\

In the following example we construct a family of rank 7 mixed states $\{\rho_{e_1,e_2,e_3}\}$ ($e_1,e_2,e_3$ are real 
parameters)  with $\rho_{e_1,e_2,e_3}=\rho_{e_1,e_2,e_3}^{PT}$ (hence PPT automatically) on $H=H_{A}^4 \otimes 
H_{B}^6$. We prove that they are entangled by Theorem 3 (thus bound entanglement) for generic parameters $e_1,e_2,e_3$ 
and parameters $(e_1,e_2,e_3)=(0,0,1)$. This family and the method used here  can be easily generalized to construct 
entangled mixed states with $\rho=\rho^{PT}$ systematically.\\

 Consider the following 4 $ 6 \times 7$ matrices\\

$$
A_1=
\left(
\begin{array}{ccccccccccc}
1&0&0&0&0&0&0\\
0&1&0&0&0&0&0\\
0&0&1&0&0&0&0\\
0&0&0&2&0&0&1\\
0&0&0&0&2&0&0\\
0&0&0&0&0&2&0
\end{array}
\right)
$$

$$
A_2=
\left(
\begin{array}{ccccccccccc}
0&1&1&-1&0&0&1\\
1&0&1&0&0&0&0\\
1&1&0&0&0&0&0\\
-1&0&0&0&1&1&0\\
0&0&0&1&0&1&0\\
0&0&0&1&1&0&0
\end{array}
\right)
$$

$$
A_3=
\left(
\begin{array}{ccccccccccc}
e_2+e_3&e_1&0&0&0&0&0\\
e_1&e_2&e_3&0&0&0&0\\
0&e_3&e_1+e_2&0&0&0&0\\
0&0&0&e_2+e_3&e_1&0&0\\
0&0&0&e_1&e_2&e_3&0\\
0&0&0&0&e_3&e_1+e_2&0
\end{array}
\right)
$$
, where $e_1,e_2,e_3$ are real parameters, and $A_4=(I_6,0)$, where $I_6$ is $6\times 6$ unit matrix.\\

Let $A$ be a $24 \times 7$ matrix with 4 blocks $A_1,A_2,A_3,A_4$ where the 24 rows correspond to the standard basis 
$\{|11 \rangle,...,|16 \rangle,...,|41 \rangle,...,|46 \rangle\}$. Let $\rho_{e_1,e_2,e_3}$ be $\frac{1}{D}A 
A^{\dagger}$ (where $D$ is a normalizing constant), a mixed state on $H$. It is easy to check that $A_i 
A_j^{\dagger}=A_j A_i^{\dagger}$, hence $\rho_{e_1,e_2,e_3}$ is invariant under partial transposition and thus PPT.\\

Let $|\psi_1 \rangle,...,|\psi_7 \rangle \in H_A^4 \otimes H_B^6$  be $7$ vectors corresponding to $7$ columns of the 
matrix $A$. It is clear that the range of $\rho_{e_1,e_2,e_3}$ is the linear span of $|\psi_1 \rangle,..,|\psi_7 
\rangle$. When $e_1=e_2=0,e_3=1$, $|\psi_2 \rangle -|\psi_3 \rangle=(|1 \rangle+|4 \rangle-|2 \rangle)\otimes 
(|2 \rangle-|3 \rangle)$. Thus there are some separable pure states in the range of $\rho_{0,0,1}$. We will show that 
$\rho_{0,0,1}$ and $\rho_{e_1,e_2,e_3}$ for generic parameters $e_1,e_2,e_3$ are entangled by our separability 
criterion Theorem 3.\\

As in the proof of Theorem 2 it is easy to compute $F=r_1A_1+r_2A_2+r_3A_3+r_4A_4$.\\                         

$$
\left(
\begin{array}{ccccccccccc}
u_1&r_2+e_1r_3&r_2&-r_2&0&0&r_2\\
r_2+e_1r_3&u_1'&r_2+e_3r_3&0&0&0&0\\
r_2&r_2+e_3r_3&u_1''&0&0&0&0\\
-r_2&0&0&u_2&r_2+e_1r_3&r_2&r_1\\
0&0&0&r_2+e_1r_3&u_2'&r_2+e_3r_3&0\\
0&0&0&r_2&r_2+e_3r_3&u_2''&0
\end{array}
\right)
(20)
$$
, where $u_1=r_1+r_4+(e_2+e_3)r_3,u_1'=r_1+r_4+e_2r_3$ ,$u_1''=r_1+r_4+(e_1+e_2)r_3$ and 
$u_2=2r_1+r_4+(e_2+e_3)r_3,u_2'=2r_1+r_4+e_2r_3$,$u_2''=2r_1+r_4+(e_1+e_2)r_3$.\\

We consider the following matrix $F'$ which is obtained by adding the 7-th  column of $F$ to the 4-th column of $F$ 
and adding $r_2/r_1$ of the 7-th column to the 1st column.\\

$$
\left(
\begin{array}{cccccccccccc}
v_1&r_2+e_1r_3&r_2&0&0&0&r_2\\
r_2+e_1r_3&u_1'&r_2+e_3r_3&0&0&0&0\\
r_2&r_2+e_3r_3&u_1''&0&0&0&0\\
0&0&0&v_2&r_2+e_1r_3&r_2&r_1\\
0&0&0&r_2+e_1r_3&u_2'&r_2+e_3r_3&0\\
0&0&0&r_2&r_2+e_3r_3&u_2''&0
\end{array}
\right)
(21)
$$
, where $v_1=r_1+r_4+(e_2+e_3)r_3+ \frac{r_2^2}{r_1}$,$v_2=3r_1+r_4+(e_2+e_3)r_3$.\\

It is clear that the determinantal varieties defined by $F$ and $F'$ are the same in the affine chart $C^3$ defined by 
$r_1\neq 0$. Consider the zero locus $Z_1$ defined by the condition that the determinants of the 2 diagonal $3\times3$ 
submtrices of the 1st $6 \times 6$ submatrix in (21)  are zero, locus $Z_2$ defined by the condition that the 1st 3 
rows in (21) are linear dependent and the locus $Z_3$ defined by the condition that the last 3 rows in (21) are linear 
dependent, it is clear that $V_A^5(\rho_{e_1,e_2,e_3}) \cap C^3$ is the sum of $Z_1,Z_2,Z_3$. We can use the affine 
coordinates $r_2'=\frac{r_2}{r_1},r_3'=\frac{r_3}{r_1},r_4'=\frac{r_4}{r_1}$ on the affine chart $C^3$ of $CP^3$ 
defined by $r_1 \neq 0$. In this affine coordinate system all $r_2,r_3,r_4$ should be replced by $r_2',r_3',r_4'$ and 
$r_1$ should be replaced by 1 in (21). Now we analysis $V_A^5(\rho_{0,0,1})$. It is clear that the following 2 planes 
$H_1=\{(r_2',r_3',r_4'): r_2'=r_4'+1\},H_2=\{(r_2',r_3',r_4'):r_2'=r_4'+2\}$ are in $V_A^5(\rho_{0,0,1}) \cap C^3$, 
since in the case $r_2'=r_4'+1$ the 2nd and the 3rd rows of (21) are linearly dependent and in the case $r_2'=r_4'+2$ 
the 5th and 6th rows of (21) are linearly dependent. The determinants of two $3 \times 3$ diagonal submatrices of the 
1st $6 \times 6$ submatrix of (21) are:\\

$$
\begin{array}{cccc}
(r_2'-r_4'-1)((r_2')^3 +(r_2')^2r_4'-(r_2')^2 +(r_4')^2\\
+ r_2'r_3'+r_2'r_4'+r_3'r_4'+r_2'+r_3'+2r_4'+1)
\end{array}
$$
and
$$
\begin{array}{cccc}
(r_2'-r_4'-2)((r_4')^2-2(r_2')^2+r_2'r_3'+r_2'r_4'+r_3'r_4'+3r_2'+2r_3'+5r_4'+6)
\end{array}
$$

Let $X_1$ and $X_2$ be zero locus of the 2nd factors of the above two determinants. It is obvious $X_1 \cap X_2$ is in 
$V_A^5(\rho_{0,0,1}) \cap C^3$, we want to show that $X_1 \cap X_2 \setminus H_1 \cup H_2$ is a curve, not a line. 
Take the ponit $P=(0,2,-1) \in X_1 \cap X_2 \cap H_1$, the tangent plane $H_3$ of $X_2$ at $P$ is defined by 
$4r_2'+r_3'+5r_4'=-3$. If $X_1 \cap X_2$ is a line around the ponit $P$, this line is contained in $H_3 \cap X_2$. 
However we can easily find that $H_3 \cap X_2$ is defined by $3(r_2')^2+2(r_4')^2+4r_2'r_4'+4r_4'=0$. This polynomial 
is irreducible and thus $H_3 \cap X_2$ is a curve around the piont $P$. Thus $X_1 \cap X_2$ is a curve around the 
point $P$. It is easy to check that $X_1 \cap X_2$ is not contained in $H_1$ around the point $P$.   This implies that 
$V_A^5(\rho_{e_1,e_2,e_3}) \cap C^3$ (actually the locus $Z_1$) contains a curve (not a line) for generic parameters 
$e_1,e_2,e_3$(including parameters $0,0,1$) from algebraic geometry. Thus if $V_A^5 (\rho_{e_1,e_2,e_3}) \cap C^3$ is 
the sum of (affine) linear subspaces, it have to contain a dimension 2 affice linear subspace $H_4$ other than $H_1$ 
and $H_2$ of the affine chart $C^3$. Thus the determinants of all $6 \times 6$ submatrices of (21) have to contain an 
(fixed) affine linear form (i.e., a degree one polynomial of $r_2',r_3',r_4'$ which may contain a constant term) other 
than $r_2'-r_4'-1$ and $r_2'-r_4'-1$ as one of their factors. This affine linear form defines that dimension 2 linear 
affine subspace $H_4$ of $C^3$. However it is easy to check this is impossible for generic parameters $e_1,e_2,e_3$ 
(including parameters $0,0,1$). We know that $V_A^5(\rho_{e_1,e_2,e_3}) \cap C^3$ cannot be the sum of (affine) linear 
subspaces of $C^3$ for generic $e_1,e_2,e_3$ (including parameters $0,0,1$).  Thus from Theorem 3 $\rho_{e_1,e_2,e_3}$ 
is entangled for generic parameters $e_1,e_2,e_3$ (including parameters $0,0,1$).\\

{\bf Theorem 12.} {\em The mixed states $\rho_{e_1,e_2,e_3}$'s, which are invariant under partial transposition,  are 
entangled for generic parameters and $(e_1,e_2,e_3)=(0,0,1)$.}\\

{\bf Remark 4.} $\rho_{0,0,1}$ is the first example of PPT entangled mixed state (thus bound entanglement) with some 
separable pure states in its range.\\

From the construction in this Example  we can see if $A_1,...,A_m$ are $m$ $n \times t$ matrices satisfying $A_i 
A_j^{\dagger}=A_j A_i^{\dagger}$, $A$ is the $m \times 1$ matrix with i-th block $A_i$ and the rows of $A$ correspond 
to the basis $|11 \rangle ,...,|1n \rangle,...,|m1 \rangle,...,|mn \rangle$ of $H_A^m \otimes H_B^n$, then the mixed 
state $\rho=\frac{1}{D}A A^{\dagger}$, where $D$ is a normalized constant, is invariant under partial transpose. It is 
not very difficult to find such matrices. For the purpose that the constructed mixed state $\rho$ is entangled (thus a 
bound entangled mixed state),  we just need that the determinantal variety $\{(r_1,...,r_m): rank (\Sigma r_i A_i) 
\leq n-1\}$ is NOT {\em linear}. We know from algebraic-geometry, it is not very  hard to find such matrices 
$A_1,...,A_m$. However as illustrated in this Example  we do need some explicit calculation to prove this point. Thus 
our separability criterion and the method used in this Example  offer a new systematic way to construct PPT bound 
entangled mixed states.\\

{\bf Acknowledgment.}  This work was supported by NNSF China grant 10171110 and NNSF China Distinguished Young Scholar 
Grant 10225106.\\

\begin{center}
REFERENCES
\end{center}

1.S.Abhyankar, Combinatorics of Young tableaux, determinantal varieties and the calculation of Hilbert functions(in 
French), Rend. Sem. Mat. Univ. Politec. Torino, 42(1984), no.3, pp65-88\\

2.A.Acin, A.Andrianov, L.Costa, E.Jane, J.I.Latorre and R.Tarrach, Generalized Schmidt decomposition and 
classification of three-qubit states, Phys. Rev. Lett. 85, 1560,2001\\

3.A.Acin, D.Bruss, M.Lewenstein and A.Sanpera, Classification mixed three-qubit states, Phys. Rev. Lett. 87, 
040401(2001)\\

4.E.Arbarello, M.Cornalba, P.A.Griffiths and J.Harris, Geometry of algebraic curves,Volume I, Springer-Verlag, 1985, 
Chapter II Determinantal Varieties\\ 

5.C.H. Bennett and P.W.Shor, Quantum Information Theory, IEEE Trans. Inform. Theory, vol.44(1998), No.6\\

6.C.H.Bennett, G.Brassard, C.Crepeau, R.Jozsa, A.Peres and W.K.Wootters, Teleporting an unknown quantum state via dual 
classical and Einstein-Podolsky-Rosen channels, Phys. Rev. Lett. 70, 1895 (1993)\\

7.C.H.Bennett, D.P.DiVincenzo, T.Mor, P.W. Shor, J.A.Smolin and T.M. Terhal, Unextendible product bases and bound 
entanglement, Phys. Rev. Lett. 82, 5385-5388, 1999\\

8.C.H.Bennett, S.Popescu, D.Rohrlich, J.A.Smolin and A.Thapliyal, Exact and asymptotic measure of multipartite 
pure-state entanglement,  Phys. Rev. A 63, 012307, 2000\\

9.C.H.Bennett, J.I.Cirac, M.S.Leifer, D.W.Leung, N.Linden, S.Popescu and G.Vidal, Optimal simulation of two-qubit 
Hamiltonians using general local operations, quant-ph/0107035, Phys.Rev. A, 66, 012305(2002)\\

10.E.Brieskorn and H.Knorrer, Plane algebraic curves, Birkhauser Boston,1981\\

11.J-L. Brylinski and R. Brylinski, Invariant polynomial functins on k qudits, Mathematics of Quantum Computation, 
edited by R.Brylinski and G. Chen CRC Press, 2002\\

12.J-L. Brylinski and R.Brylinski, Universal quantum gates, Mathematics of Quantum Computation, edited by R.Brylinski 
and G. Chen, CRC Press, 2002 \\

13.H. Chen, Necessary conditions for efficient simulation of Hamiltonians using local unitary operations, 
quant-ph/0109115, Quantum Information and Computation Vol.3,No.3 (2003), pp249-257\\

14.H. Chen, Schmidt numbers of low rank bipartite mixed states, quant-ph/0211052, Phys. Rev. A, Vol.67(2003),062301\\

15.C.De Concini, D.Eisenbud and C.Procesi, Young diagrams and determinantal varieties, Invent. Math., 56(1980), no.2, 
pp129-165\\

16.A.Ekert and R.Jozsa, Quantum algorithms: entanglement enhanced information processing, Phil. Trans. Roy. Soc. 
(Lond.), 1998\\

17.A.Ekert, Quantum cryptography based on Bell's theorem, Phys Rev. Lett. 67, 661(1991)\\

18.A.Einstein, B.Podolsky and N.Rosen, Can quantum mechanical description of physical reality be considered complete, 
Phys. Rev. 47, 777(1935)\\

19.D.Eisenbud, J.Koh and M.Stillman, Determinantal equations for curves of high degree, Amer.J.Math., 
vol.110(1988),pp513-539\\

20.D.Eisenbud, Linear sections of determinantal varieties, Amer.J.Math., Vol.110(1988), pp.541-575\\

21.D.M.Greenberger, M.Horne and A.Zeilinger, Bell's Theorem, Quantum Theory, and Conceptions of the Universe, edited 
by M.Kafatos (Kluwer Dordrecht)\\

22.P.A.Griffiths, Infinitesimal variations of Hodges structure III: Determinantal varieties and the infinitesimal 
invariants of normal functions, Compositio Math., 50(1983), no.2-3, pp267-324\\

23.J.Harris, Algebraic geometry, A first Course, Gradute Texts in Mathematics, 133, Springer-Verlag, 1992,
  especially Lecture 9  Determinantal Varieties\\

24.M.Horodecki, P.Horodecki and R.Horodecki, Separability of mixed states: Necessary and sufficient conditions, Phys. 
Lett. A 223, 8 (1996)\\

25.P.Horodecki,Separability criterion and inseparable mixed states with positive partial transpose, Phys. Lett. A 232 
333(1997)\\

26.M.Horodecki, P.Horodecki and R.Horodecki,  Mixed state entanglement and distillation: is there ``bound'' 
entanglement in nature, Phys. Rev. Lett. 80, 5239(1998)\\

27.M.Horodecki, P.Horodecki and R.Horodecki, Separability of n-particle mixed states: 
necessary and sufficient conditios in terms of linear maps, quant-ph/0006071, Phys. Lett. A, 283(2001)\\

28.M.Horodecki , P.Horodecki and R. Horodecki, Mixed state entanglement and quantum communication, in ``Quantum 
Infomation---Basic concepts and experiments'', Eds, G.Alber and M.Wiener (Springer Berlin)\\

29.P.Horodecki, M.Lewenstein, G.Vidal and J.I. Cirac,Operational criterion and constructive checks for the 
separability of low rank density matrices, Phys. Rev. A 62, 032310 (2000)\\

30.B.Kiraus, J.I.Cirac, S.Karnas and M.Lewenstein, Separability in 2xN composite quantum systems, Phys. Rev. A 61, 
062302 (2000)\\

31.L.H.Kauffman and S.J.Jr.Lomonaco, Quantum entanglement and topological entanglement, New Journal of Physics, 4 
,73(2002)\\

32.M. Lewenstein, D Bruss, J.I.Cirac, B.Krus, M.Kus, J. Samsonowitz , A.Sanpera and R.Tarrach, Separability and 
distillablity in composite quantum systems--a premier, J.Mod. Optics, 47, 2481(2000)\\

33.N.Linden and S.Popescu,On multi-particle entanglement, Fortsch. Phys.,46, 567(1998)\\

34.N.Linden, S.Popescu and A.Sudbery, Non-local parameters of multi-particle density matrices, Phys. Rev. Lett, 
83(1999), 243-247\\

35.D.Meyer and N.R.Wallach, Global entanglement in multiparticle systems, J. Math. Phys., 4273-4278, no.9, 43(2002)\\

36.M.A.Nielsen, Conditions for a class of entanglement transformations, Phys. Rev. Lett. 83, 436 (1999)\\

37.M.A.Nielsen and J.Kempe, Separable states are more disordered globally than locally, Phys. Rev. Lett., 86, 5184 
(2001)\\

38.A.Peres, Separability criterion for density matrices, Phys. Rev. Lett. 77,1413-1415(1996)\\

39.J.Preskill, Physics 229: Advanced mathematical methods of Physics--Quantum Computation and Information (California 
Institute of Technology, Pasadena,CA,1998), http://www.theory.caltech.edu/people/preskill/ph229/\\

40.J.A.Smolin, Four-party unlockable bound-entangled states, Phys. Rev. A 63, 032306(2001)\\

41.B.Terhal and P.Horodecki, Schmidt number for density matrices,  Phys. Rev. A61, 040301(2000)\\

42.B.M.Terhal, Detecting quantum entanglements, quant-ph/0101032, Theoret. Comput. Sci., 287(2002),no.1 \\

43.N.R.Wallach, An unentangled Gleason's theorem, Contemp. Math. 305 (2002), AMS, Providence, RI\\

44.R.F.Werner and M.M.Wolf, Bell's inequality for states with positive partial transpose, Phys. Rev. A 61, 062102 
(2000)\\

\end{document}